\documentclass[lettersize,journal]{IEEEtran}
\usepackage{amsmath,amsfonts}
\usepackage{amssymb}
\usepackage{algorithmic}
\usepackage{algorithm}
\usepackage{array}
\usepackage[caption=false,font=normalsize,labelfont=sf,textfont=sf]{subfig}
\usepackage{textcomp}
\usepackage{stfloats}
\usepackage{url}
\usepackage{verbatim}
\usepackage{graphicx}
\newcommand{\compactbio}[3]{%
\par\addvspace{0.8\baselineskip}%
\noindent
\begin{minipage}{\columnwidth}
\begin{minipage}[t]{0.20\columnwidth}
\vspace{0pt}
\IfFileExists{#1}
{\includegraphics[width=\linewidth,height=1.05in,keepaspectratio]{#1}}
{\fbox{\rule{0pt}{1.05in}\rule{0.75in}{0pt}}}
\end{minipage}
\hfill
\begin{minipage}[t]{0.76\columnwidth}
\vspace{0pt}
\scriptsize \textbf{#2} #3
\end{minipage}
\end{minipage}
}
\usepackage{cite}
\usepackage[colorlinks=true,linkcolor=blue,citecolor=blue,urlcolor=blue]{hyperref}
\AtBeginDocument{\setlength{\abovedisplayskip}{3pt plus 1pt minus 1pt}\setlength{\belowdisplayskip}{3pt plus 1pt minus 1pt}\setlength{\abovedisplayshortskip}{2pt plus 1pt minus 1pt}\setlength{\belowdisplayshortskip}{2pt plus 1pt minus 1pt}}

\begin{document}

\title{RQ-SAFE: Coupled Request--Resource Scheduling for Online Edge SFC-DAGs}

\author{Shengdong Gu, Hongyuan Wan, and Taixin Li
\thanks{Shengdong Gu is with the School of Data Science and Artificial Intelligence, Dongbei University of Finance and Economics, Dalian, China.}
\thanks{Hongyuan Wan is with the School of Management Science and Engineering, Dongbei University of Finance and Economics, Dalian, China.}
\thanks{Taixin Li is with the School of Data Science and Artificial Intelligence, Dongbei University of Finance and Economics, Dalian, China.}
\thanks{Corresponding author: Taixin Li.}}

\markboth{IEEE Transactions on Network and Service Management,~Vol.~XX, No.~X, Month~2026}%
{Gu \MakeLowercase{\textit{et al.}}: Coupled Request--Resource Scheduling for Edge SFC-DAGs}

\maketitle

\begin{abstract}
Intent-driven edge services allow multiple virtual network function (VNF) segments in a service function chain directed acyclic graph (SFC-DAG) to be locally reordered without changing service semantics, creating richer request-side orchestration freedom. Existing orchestration methods mainly optimize VNF placement, routing, or queue-aware scheduling for a predetermined service order; they do not fully exploit this freedom or couple it with runtime resource scheduling. This paper presents RQ-SAFE, a request--resource coupled scheduling framework for online edge SFC-DAG orchestration with checked commitment. RQ-SAFE evaluates each feasible local order by previewing its resource-side consequences on the current edge infrastructure, and uses the retained order to guide VNF instance selection and path construction. Queue state is used throughout the decision process to evaluate local orders, rank per-VNF candidates, and perform final queue-aware quality-of-service (QoS) validation. A profile-aware learning-assisted re-ranker balances request-side QoS objectives and resource-side load objectives by refining retained top-$K$ candidates. On matched edge SFC-DAG workloads, RQ-SAFE achieves comparable QoS-compliant service outcomes to graph-aware baselines while improving resource balance. Relative to the graph neural network (GNN)-based GNN-DAG-Score baseline on public-mixed workloads, it reduces central processing unit (CPU) imbalance by $6.1\%$ and peak CPU by $2.3\%$, with limited additional control-plane decision time. Ablation results show that enabling local-order flexibility and queue awareness together improves QoS by $4.53$ percentage points over disabling both factors, with a $3.83$ percentage-point positive interaction between the two factors. Overall, RQ-SAFE offers a practical request--resource coupling paradigm for exploiting orchestration freedom in intent-aware edge SFC-DAG services.
\end{abstract}


\newpage
\section{Introduction}

\IEEEPARstart{E}{dge} computing, software-defined networking (SDN), and network function virtualization (NFV) allow service functions to run on shared infrastructure instead of dedicated appliances~\cite{kreutz2015sdn,etsi2014nfv}. Virtual network functions (VNFs) are commonly connected as service function chains (SFCs) for security inspection, traffic filtering, encryption, caching, and forwarding~\cite{halpern2015sfc,mijumbi2016nfv,bhamare2016sfc}. At the edge, orchestration must operate under limited node capacity, heterogeneous link delay, and time-varying demand~\cite{shi2016edge,taleb2017mec,premsankar2018edgeiot}. Each active VNF instance also has running tasks, waiting requests, and finite service capability. A placement that looks feasible from static processing and path delay may therefore violate its quality-of-service (QoS) budget after the traffic reaches a congested instance~\cite{chen2023queueaware}.

Intent-driven management is also making service requests more flexible. A service objective, security rule, or QoS target can be specified while part of the VNF order remains open~\cite{ibn_survey,ibn_kg,ibn_vec}. The resulting service may contain branches, merges, and functions whose relative order is only partially constrained. It is naturally represented as an SFC directed acyclic graph (SFC-DAG) or VNF forwarding graph~\cite{mehraghdam2014specifying,xie2022vnffg}. This representation exposes request-side orchestration freedom: several local execution orders can preserve service semantics but lead to different VNF instances, paths, queue pressure, and load-distribution outcomes.

This freedom cannot be used effectively without observing the current resource state. A local rewrite that is beneficial under one service-pool snapshot can become inferior when queues or node loads change. Likewise, reusing a warm instance can reduce activation delay but consume the remaining QoS margin if that instance is busy. The orchestration problem therefore goes beyond choosing an order and then placing VNFs. Request-side order choices should be evaluated through their resource-side consequences, and resource-side placement should remain conditioned on the selected request structure.

Recent placement and protection methods improve resource allocation for supplied SFCs~\cite{huang2025deviceless,zheng2025protection}. Parallelized-SFC studies further model dynamic resource demand and uncertain traffic~\cite{zhang2025dynamic,zhang2025robust}. Queue-aware schedulers make congestion visible during routing or placement, and learning-based methods can absorb rich runtime features. These lines provide important building blocks. RQ-SAFE addresses a missing decision: how a bounded set of semantics-preserving local reordering actions should be used under the current queue, path, service-pool, and load state, and how the selected order should guide detailed instance construction.

This paper presents \emph{RQ-SAFE}, a request--resource coupled framework for joint online scheduling of edge SFC-DAGs. RQ-SAFE extracts a compact set of semantics-preserving local-order actions and evaluates each action through a placement-aware preview. This preview links request completion and QoS requirements with queue, path, service-pool, and load conditions before the order is retained. Detailed placement then proceeds under the retained order, and candidate scores are recomputed as the partial plan evolves. Learning-assisted re-ranking is used as a bounded refinement inside the retained top-$K$ candidate neighborhood, while action choices, fallbacks, validation checks, and outcomes remain traceable.

The main contributions are as follows.
\begin{itemize}
    \setlength{\itemsep}{1pt}
    \item We formulate online edge SFC-DAG orchestration as request--resource joint scheduling. A request-side order action is evaluated by the placement result it would produce, and the selected order then shapes the candidate instances and paths considered on the resource side.
    \item We develop a semantics-preserving local reordering mechanism and a recursive order-aware candidate model. The model jointly considers QoS margin, queue pressure, path delay, service reuse, projected node load, hotspot risk, and candidate availability.
    \item We make queue state the runtime coupling signal between request-side order flexibility and resource-side placement. Queue-aware feedback guides local-order evaluation, per-VNF candidate ranking, and final QoS validation with queueing delay. The profile-aware learning-assisted re-ranker then adjusts the tradeoff between QoS and load within retained top-$K$ candidates, while constrained selection and request-level records keep the decision path traceable.
\end{itemize}

The remainder of this paper is organized as follows. Section~II positions the work. Section~III defines the request--resource coupling model. Section~IV presents RQ-SAFE. Section~V reports the evaluation, and Section~VI concludes the paper.

\section{Related Work}

\textit{SFC placement, routing, and protection.}
Classical SFC orchestration maps VNFs to physical nodes and routes service traffic through the selected instances under node, link, and delay constraints~\cite{herrera2016nfv,bari2016orchestrating}. Edge-oriented methods further consider service deadlines, heterogeneous edge devices, instance reuse, and protection requirements~\cite{nguyen2020placement,jin2020latency,wang2021adaptive,huang2025deviceless,zheng2025protection}. These studies form the resource-feasibility foundation for online SFC deployment. They also show that placement quality depends on multiple runtime factors rather than on CPU capacity alone. However, their decision input is usually a chain or service graph that has already been specified. The order in which functions should be executed is not treated as a runtime decision coupled with placement. RQ-SAFE keeps the familiar placement and path-construction view, but adds an earlier decision: which legal local order should be used before detailed instance selection begins.

This distinction matters for edge environments because the placement decision is often made online and under partial resource visibility. A request can be accepted, delayed, or rejected according to the instance and path state that exists at the arrival time. If the request order is treated as immutable, a scheduler can still improve path selection or load balancing, but it cannot ask whether an equivalent local order would avoid a congested instance or a hot node. RQ-SAFE therefore keeps the standard VNF-to-node and virtual-edge-to-path mapping interface, while moving one request-interpretation decision into the online loop.

\textit{SFC-DAGs and parallelized deployment.}
VNF forwarding graphs and SFC-DAGs extend linear chains by representing branches, merges, partial ordering, and parallel processing~\cite{mehraghdam2014specifying,quang2019deep,houidi2020dynamic,cai2021appm,zhang2024parallelized}. Recent studies optimize parallelized SFCs under dynamic resource requests, uncertain traffic arrival rates, and multi-domain migration~\cite{zhang2025dynamic,zhang2025robust,zhang2025multidomain}. These works are important because they move service descriptions beyond a single fixed chain. Their main concern, however, is how a supplied graph or parallel structure should be embedded, migrated, or protected. RQ-SAFE addresses a complementary question. It does not enumerate all topological orders or serialize independent branches. Instead, it exposes a compact set of semantics-preserving local reordering actions and asks whether any of them becomes useful under the current service-pool, path, queue, and load state.

The difference is not only the size of the graph. In a forwarding graph, two services may be unordered because they are independent, or because their order is semantically interchangeable in a local context. Treating both cases as arbitrary topological choices can create a large and hard-to-audit action space. Treating the graph as fixed loses the opportunity to exploit the interchangeable segment. RQ-SAFE takes a middle position: it records branch and merge structure, but turns only declared and structurally guarded serial pairs into runtime local-order actions.

\textit{Queue-aware scheduling and QoS control.}
Queue-aware routing, fair scheduling, rate control, stochastic scheduling, and online SFC control use queue length, waiting time, or service rate to capture runtime congestion~\cite{pei2021resource,zu2021fair,yang2023online,li2023stochastic,he2024efficient,chen2023queueaware}. These works motivate our use of instance-level queue state and QoS validation with queueing delay. In representative formulations, queue state changes resource-side routing or placement after the service order is given. This is valuable but still leaves request-side order freedom outside the queue-aware loop. RQ-SAFE uses queue-aware provisional placement to evaluate legal local orders themselves. Thus, queue state is not only a delay correction after placement; it is a signal that changes how request-side orchestration freedom is used.

This placement of queue state also affects how QoS is interpreted. A delay budget can be satisfied by a static path-delay estimate and still fail after deployment if the chosen instance has accumulated work. Conversely, a path with slightly higher propagation delay can be preferable when it reaches a low-waiting warm instance. RQ-SAFE therefore uses queue state before, during, and after candidate construction: before placement to compare legal local orders, during placement to rank instances, and after placement to validate the complete end-to-end delay with queueing.

\textit{Learning-based and lifecycle-aware orchestration.}
Learning has been used for VNF placement, graph embedding, dynamic SFC deployment, and multiobjective coordination~\cite{xiao2019nfvdeep,solozabal2020drl,fu2020dynamic,chen2021drlqor,schneider2021self,bi2021multi,zhou2025multiobjective}. Such methods can absorb graph and runtime features, but an end-to-end policy over orders, instances, paths, and access decisions can be difficult to inspect and validate online. Recent lifecycle-aware work also shows that activation, reconfiguration, and reuse state should remain visible to placement logic~\cite{giarre2026ripple}. RQ-SAFE therefore places learning at a narrower boundary. Semantic action generation, placement-aware preview, feasibility checks, and fallback remain explicit. The learning-assisted component only refines the order of retained instance candidates inside a filtered neighborhood.

This design also separates two forms of adaptivity. The first is structural adaptivity: the orchestrator may choose a different legal local order when runtime state changes. The second is candidate adaptivity: for the retained order, the orchestrator may prefer a different instance or path as queues, reuse opportunities, and node pressure evolve. A broad learned policy could merge these forms into one action, but then the reason for a decision is harder to isolate. RQ-SAFE keeps them as explicit interfaces and uses learning only where a bounded candidate neighborhood has already been constructed.

\textit{Position of RQ-SAFE.}
The above lines are complementary rather than competing. Placement and routing provide the resource substrate, SFC-DAG studies provide the service-graph representation, queue-aware methods expose runtime congestion, and learning-based orchestration offers candidate refinement. The gap is their decision boundary: existing methods typically optimize a supplied order or learn a broad action space. RQ-SAFE focuses on the missing middle layer between request semantics and resource placement. It evaluates a small set of legal local orders through live placement preview and then lets the retained order guide recursive instance and path construction. Table~\ref{tab:positioning} summarizes this positioning.

\begin{table*}[!t]
\caption{Positioning of RQ-SAFE Relative to Representative Orchestration Lines}
\label{tab:positioning}
\centering
\scriptsize
\setlength{\tabcolsep}{3.5pt}
\renewcommand{\arraystretch}{1.05}
\begin{tabular}{p{0.17\textwidth}p{0.23\textwidth}p{0.25\textwidth}p{0.29\textwidth}}
\hline
\textbf{Approach line} & \textbf{Request-side treatment} & \textbf{Runtime-state treatment} & \textbf{Decision boundary relative to RQ-SAFE} \\
\hline
Placement, routing, and protection~\cite{huang2025deviceless,zheng2025protection} & Fixed chain or supplied service graph & Resource, path, delay, heterogeneity, or protection costs & Optimizes deployment of the supplied request; local-order freedom is handled outside the runtime decision \\
Parallelized SFC/SFC-DAG~\cite{cai2021appm,zhang2025dynamic,zhang2025robust,zhang2025multidomain} & Rich branch/merge or parallel structure & Dynamic demand, uncertain traffic, migration, and placement state & Optimizes a supplied parallel structure rather than a bounded semantic rewrite chosen through live placement preview \\
Queue-aware scheduling~\cite{chen2023queueaware,he2024efficient} & Service order is generally given & Queue length, waiting time, service rate, and QoS are explicit & Queue state changes routing or placement; RQ-SAFE also lets it change the use of legal order freedom \\
Learning-based orchestration~\cite{xie2022vnffg,zhou2025multiobjective} & Graph or placement actions may be learned jointly & Runtime features enter a learned score or policy & RQ-SAFE separates semantic actions and placement-aware preview from retained-candidate re-ranking \\
RQ-SAFE & Bounded linear or branch-local semantic rewrites & Order-aware completion, queue, path, service-pool, and load state & Closes an action-level preview loop and a stage-level placement loop before queue-aware checked commitment \\
\hline
\end{tabular}
\end{table*}

\section{Request--Resource Coupling Model}
\label{sec:system_model}

This section retains only the background needed to define the paper-specific decision interface. Standard SFC mapping and routing constraints are summarized in the appendix. The focus here is how partial-order freedom becomes an actionable online variable and how the request and resource sides are joined into one recursive decision process. Table~\ref{tab:notations} lists the symbols used in this section and in the online orchestration algorithm.

\begin{table}[!t]
\caption{Key Notations}
\label{tab:notations}
\centering
\scriptsize
\setlength{\tabcolsep}{3pt}
\renewcommand{\arraystretch}{0.98}
\begin{tabular}{@{}>{\raggedright\arraybackslash}p{0.28\columnwidth}>{\raggedright\arraybackslash}p{0.66\columnwidth}@{}}
\hline
\textbf{Notation} & \textbf{Description} \\
\hline
$\mathcal S(t)$ & Current service pools, queues, node loads, links, and active requests. \\
$G_i=(\mathcal V_i,\mathcal E_i)$ & SFC-DAG of request $r_i$. \\
$\mathcal C_i$ & Semantically and structurally admissible local reordering pairs. \\
$\mathcal A_i^{\rm ord}$ & Bounded request-side local-order action set. \\
$\pi_i^a$ & Orchestration order induced by action $a$. \\
$\mathcal P_{i,<v}^{a}$ & Partial plan before placing VNF $v$ under action $a$. \\
$\widetilde{\mathcal A}_{i,v}^{a}(t)$ & Filtered instances for $v$ under the partial plan. \\
$c^{\rm req},c^{\rm res},c^{\rm cpl}$ & Request-side, resource-side, and coupled candidate costs. \\
$\widehat{\mathcal P}_i(a)$ & Provisional plan produced for order action $a$. \\
$\mathbf g_i(a,t)$ & Placement-aware preview vector of order action $a$. \\
$\mathcal T^K_{i,v}$ & Retained top-$K$ instance candidates. \\
$\widehat D_i^{\rm e2e},D_i^{\max}$ & Estimated delay including queueing and request budget. \\
$\mathcal H_i$ & Per-request trace of actions, candidates, checks, and result. \\
\hline
\end{tabular}
\end{table}

\subsection{System and Request State}

\begin{figure*}[!t]
  \centering
  \includegraphics[width=0.95\textwidth]{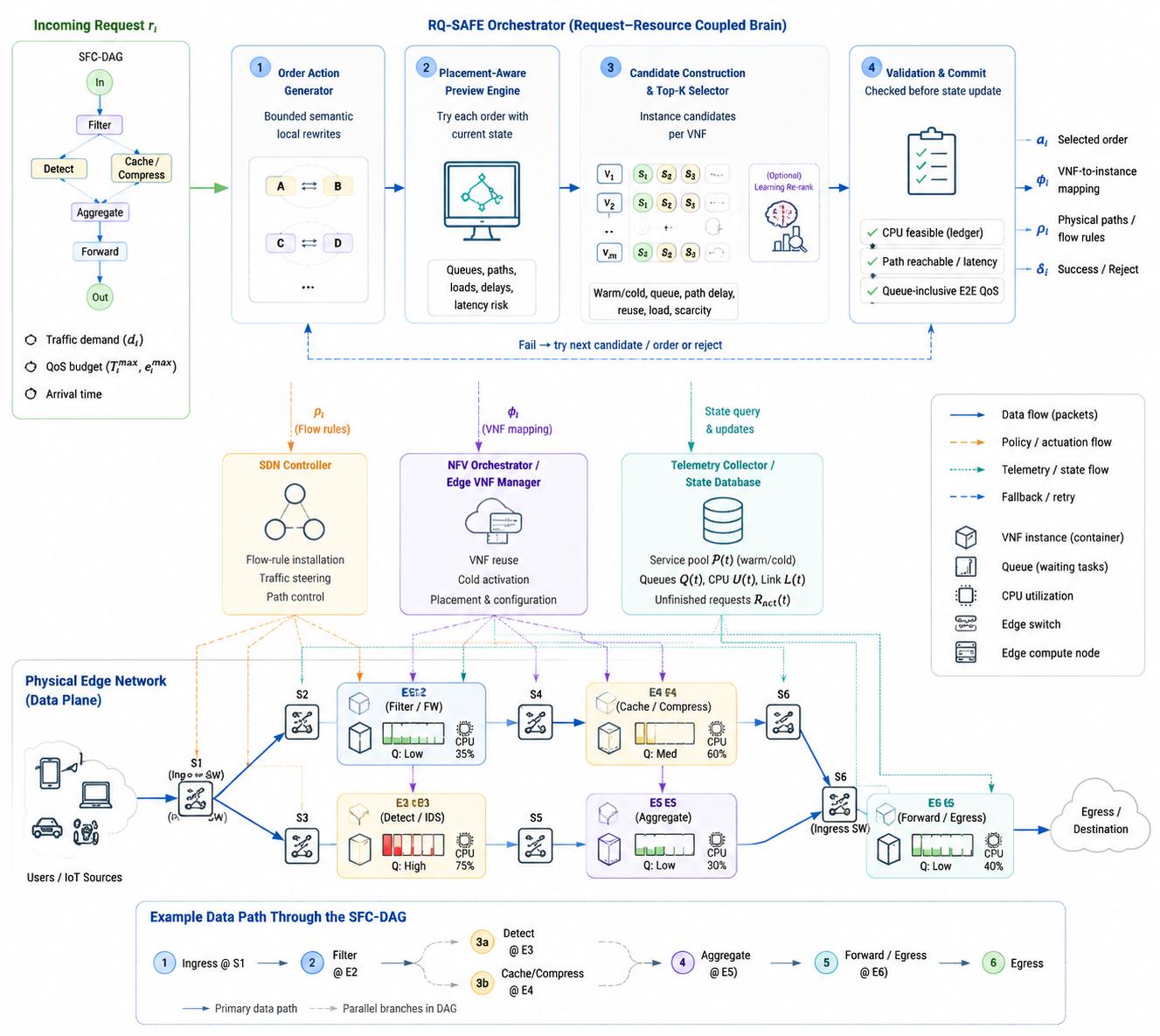}
  \caption{System-level view of RQ-SAFE in an SDN/NFV-enabled edge deployment. Solid arrows denote packet traffic steered through selected VNF instances, dashed arrows denote policy and actuation commands from the orchestrator through SDN/NFV controllers, and dotted arrows denote telemetry feedback. RQ-SAFE couples bounded request-side local reordering actions with queue/resource/path-aware placement preview, candidate construction, learning-assisted retained-candidate re-ranking, and queue-aware validation before checked state update.}
  \label{fig:overall_system}
\end{figure*}

Fig.~\ref{fig:overall_system} gives the deployment view used by the model. The orchestrator sits above SDN/NFV control and edge data-plane components, receives telemetry from queues, links, service pools, and node loads, and commits only the checked placement and path plan that will steer packet traffic through selected VNF instances.

The edge infrastructure is a directed graph $G_E=(\mathcal N,\mathcal L)$. At time $t$, the orchestrator observes
\begin{equation}
\mathcal S(t)=\langle \mathcal P(t),\mathbf Q(t),\mathbf U(t),\mathbf L(t),\mathcal R^{\rm act}(t)\rangle,
\label{eq:system_state}
\end{equation}
where $\mathcal P(t)$ is the warm/cold VNF service pool, $\mathbf Q(t)$ is the instance-level queue state, $\mathbf U(t)$ is node utilization, $\mathbf L(t)$ is link state, and $\mathcal R^{\rm act}(t)$ contains unfinished requests. Instance $s$ has VNF type $k(s)$, waiting estimate $\widehat W_s(t)$, and hosting node $n(s)$. Mapping VNF $v$ to $s$ adds an effective CPU increment $\Delta r^{\rm cpu}_{v,s}$ to the request-local ledger. This increment distinguishes cold activation from conservative warm-instance reuse.

Request $r_i$ contains an SFC-DAG $G_i=(\mathcal V_i,\mathcal E_i)$, traffic demand, QoS class, priority, and end-to-end delay budget $D_i^{\max}$. Its online decision is

\begin{equation}
X_i(t)=\langle a_i,\phi_i,\rho_i,\delta_i\rangle,
\label{eq:joint_decision}
\end{equation}

where $a_i$ is a legal local-order action, $\phi_i$ maps VNFs to instances, $\rho_i$ maps dependency edges to paths, and $\delta_i$ is the final success/reject decision. The key difference from a fixed-order formulation is that $a_i$ is evaluated through the placement process rather than fixed before it.

\subsection{Request-Side Freedom}

At runtime, RQ-SAFE uses a compact set of request-side actions rather than all topological orders. Let $A_i^{\rm loc}(u,v)=1$ indicate that $(u,v)$ is a local pair eligible for structural inspection. In a linear chain, it is an internal adjacent pair. In a branch or general DAG, it is a directed serial pair on one branch whose two endpoints have unique local predecessor/successor context; merge and tail nodes are excluded. A pair is exposed only when its VNF types are declared semantically compatible and the local rewrite preserves the surrounding graph. We define
\begin{small}
\begin{equation}
\begin{aligned}
\mathcal C_i=\{(u,v):\;&A_i^{\rm loc}(u,v)=1,\ \Psi_{\tau_i}(k(u),k(v))=1,\\
&\Gamma_i(u,v)=1\}.
\end{aligned}
\label{eq:commutable_set}
\end{equation}
\end{small}
where $\Psi_{\tau_i}$ is the request-family compatibility rule and $\Gamma_i$ is the structural guard. Unknown or stateful combinations default to non-commutable.

Starting from a valid base orchestration representation $\pi_i^0$, RQ-SAFE exposes the no-change action and at most one local rewrite per admitted pair:
\begin{small}
\begin{equation}
\begin{aligned}
\mathcal A_i^{\rm ord}&=\{a_i^0\}\cup
\{a_{i,u,v}^{\rm swap}:(u,v)\in\mathcal C_i\},
|\mathcal A_i^{\rm ord}|\le 1+|\mathcal C_i|.
\end{aligned}
\label{eq:bounded_order_actions}
\end{equation}
\end{small}
For a linear request, the action exchanges the adjacent internal pair. For an admissible branch-local serial pair, it rewrites only the local segment $p\!\rightarrow\!u\!\rightarrow\!v\!\rightarrow\!s$ as $p\!\rightarrow\!v\!\rightarrow\!u\!\rightarrow\!s$. All other edges are retained. Independent siblings remain parallel and are not converted into a semantic order action. Eq.~\eqref{eq:bounded_order_actions} therefore converts abstract partial-order freedom into a small and auditable online set without enumerating all topological orders.

\subsection{Order-Aware Joint Candidate Model}

For action $a$ and partial plan $\mathcal P_{i,<v}^{a}$, type compatibility, service availability, request-local available CPU capacity, and configured path construction form the candidate set
\begin{small}
\begin{equation}
\begin{aligned}
\widetilde{\mathcal A}_{i,v}^{a}(t)=\{s:\;&k(s)=k(v),s\text{ is available and locally feasible}\}.
\end{aligned}
\label{eq:filtered_candidates}
\end{equation}
\end{small}
The set and the value of a candidate both depend on $a$. A different order changes the predecessor nodes already selected, the path entering $v$, the remaining VNFs, the available service pools, and the QoS slack at this stage.

The request side asks whether a candidate preserves the service's completion capability. We summarize its completion-oriented cost as
\begin{equation}
\begin{aligned}
c^{\rm req}_{i,v,s}(a,t)=&\lambda_p\widehat d^{\rm svc}_{v,s}
+\lambda_l d^{\rm path}_{i,v,s}(a,t)\\
&+\lambda_{\tau}d^{\rm ct}_{i,v,s}(a,t)
+\lambda_f d^{\rm future}_{i,v,s}(a,t),
\end{aligned}
\label{eq:req_cost}
\end{equation}
where the terms describe processing/activation completion, predecessor-to-candidate path delay, critical/merge/tail exposure, and the risk of consuming a scarce candidate needed later. The resource side asks whether the same choice avoids current congestion and future concentration:
\begin{equation}
\begin{aligned}
c^{\rm res}_{i,v,s}(a,t)=&\lambda_q\widehat W_s(t)
+\lambda_u p^{\rm cpu}_{v,s}(t)
+\lambda_h H_{i,v,s}(a,t)\\
&+\lambda_b B_{i,v,s}(a,t)-\lambda_w R_{v,s}(t),
\end{aligned}
\label{eq:res_cost}
\end{equation}
where $H$ and $B$ represent hotspot and branch-concentration pressure, and $R$ is a bounded warm/service-pool reuse benefit. Projected CPU pressure is
\begin{equation}
p^{\rm cpu}_{v,s}(t)=
\frac{U^{\rm cpu}_{n(s)}(t)+\Delta r^{\rm cpu}_{v,s}}
{C^{\rm cpu}_{n(s)}}.
\label{eq:projected_cpu}
\end{equation}
Queue pressure remains separate from CPU pressure because a node can have sufficient available CPU capacity while its matching service instance is busy.

The two sides enter one order-aware preference:
\begin{equation}
c^{\rm cpl}_{i,v,s}(a,t;p)=
\beta_p^{\rm req}c^{\rm req}_{i,v,s}(a,t)
+\beta_p^{\rm res}c^{\rm res}_{i,v,s}(a,t),
\label{eq:coupled_cost}
\end{equation}
where deployment profile $p$ sets the relative emphasis on completion/QoS and load pressure. Because the components in Eqs.~\eqref{eq:req_cost} and~\eqref{eq:res_cost} have different physical units, each raw term is first converted by a fixed bounded transform into a dimensionless score. The scaling constants are taken from the configuration or validation split and remain fixed for the held-out test runs.

The profile coefficients $\beta_p^{\rm req}$ and $\beta_p^{\rm res}$ are nonnegative and are reported on the normalized scale $\beta_p^{\rm req}+\beta_p^{\rm res}=1$, which preserves the induced ordering. A conservative profile assigns more emphasis to completion slack, critical stages, and tail exposure, whereas a throughput-oriented profile tolerates greater projected pressure when reusable service capacity is available. The profile changes candidate preference while preserving the semantic action set, feasibility filters, and recursive coupling structure.

The implementation realizes this preference through bounded base, critical-path, tail/future, and resource/reuse score groups. Eq.~\eqref{eq:coupled_cost} is their compact modeling interface rather than a claim of global scalarized optimality.

The joint process is recursive. Under action $a$, VNF $v$ is considered using the current partial plan, and a retained candidate is chosen from the coupled ranking:

\begin{small}
\begin{equation}
\begin{aligned}
\mathcal T^K_{i,v}(a,t)&=\operatorname{TopK}_{s\in\widetilde{\mathcal A}_{i,v}^{a}(t)}
\{-c^{\rm cpl}_{i,v,s}(a,t;p)\},
\mathcal P_{i,\le v}^{a}=\mathcal P_{i,<v}^{a}\oplus(v,s_v).
\end{aligned}
\label{eq:recursive_joint_placement}
\end{equation}
\end{small}

After the tentative assignment, predecessor locations, path costs, projected node pressure, reuse opportunities, and future scarcity all change. They are recomputed for the next VNF. The resource side therefore feeds back after every stage of the retained order rather than assigning one fixed score to the whole request.

\subsection{Placement-Aware Preview and Coupling Loop}

A legal action is useful only if it leads to a better provisional plan under the current state. RQ-SAFE applies the recursive process in Eq.~\eqref{eq:recursive_joint_placement} to build
\begin{equation}
\widehat{\mathcal P}_i(a)=\mathcal G(\pi_i^a,\mathcal S(t)),
\label{eq:provisional_plan}
\end{equation}
where $\mathcal G$ denotes the lightweight placement preview. Its observable consequences are summarized as
\begin{equation}
\begin{aligned}
\mathbf g_i(a,t)=\langle &\chi_i(a),\widehat D_i^{\rm proc}(a),
\widehat D_i^{\rm queue}(a),\\
&\widehat D_i^{\rm net}(a),\widehat D_i^{\rm e2e}(a),
D_i^{\max}-\widehat D_i^{\rm e2e}(a)\rangle,
\end{aligned}
\label{eq:order_preview}
\end{equation}
where $\chi_i(a)$ indicates whether the provisional placement succeeds. The action comparison rule first favors feasibility and then lower completion cost including queueing. It retains the original order or a better local rewrite according to these runtime consequences.

Eqs.~\eqref{eq:recursive_joint_placement}--\eqref{eq:order_preview} show how the two sides are connected. At the action level, a request-side rewrite induces a full provisional placement, and the resource state decides whether that rewrite is useful. At the stage level, the retained order determines the predecessor context, while every tentative resource choice updates the partial plan used by the next stage. This two-way dependency is the joint scheduling mechanism. It is stronger than placing VNFs after a fixed order, and more controlled than learning one unrestricted order--placement action.

A concrete branch-local example illustrates the two-way dependence. Consider two legal alternatives, $p\!\rightarrow\!\mathrm{Detect}\!\rightarrow\!\mathrm{Cache}\!\rightarrow\!s$ and $p\!\rightarrow\!\mathrm{Cache}\!\rightarrow\!\mathrm{Detect}\!\rightarrow\!s$. Suppose that the nearest detection instance currently has a long queue, while a low-waiting warm cache instance is available on another edge node. Under the base order, placing Detect first consumes part of the delay budget and fixes the predecessor node and path used to evaluate Cache. Under the rewrite, Cache is placed first, so its selected node changes the path, queue pressure, and feasible detection candidates seen by the next stage. After this tentative assignment, Eq.~\eqref{eq:recursive_joint_placement} updates the partial plan and recomputes all candidate terms for the second VNF. A later queue-state change can reverse the preferred order. Thus, the two orders do not receive fixed priorities; they are compared through the state transitions induced by their provisional placements.

\subsection{Queue Awareness Throughout the Online Model}

Queue information enters at three points. First, $\widehat W_s(t)$ changes the provisional placement and hence the comparison among legal actions. Second, it changes the per-VNF candidate ranking in Eq.~\eqref{eq:res_cost}. Third, the complete plan is evaluated with end-to-end delay including queueing. For VNF $v$ mapped to $s=\phi_i(v)$,
\begin{equation}
\widehat T_i^{\rm finish}(v)=\widehat T_i^{\rm start}(v)
+\widehat d^{\rm proc}_{v,s}+\widehat W_s(t)+D_s^{\rm start},
\label{eq:finish_time}
\end{equation}
where $D_s^{\rm start}$ is zero for an active instance and is the configured activation delay for a cold candidate. The start time follows the DAG precedence relation:
\begin{equation}
\widehat T_i^{\rm start}(v)=
\begin{cases}
\alpha_i, & \operatorname{Pred}(v)=\emptyset,\\
\displaystyle\max_{u\in\operatorname{Pred}(v)}
\{\widehat T_i^{\rm finish}(u)+D^{\rm net}_{i,u,v}\}, & \text{otherwise}.
\end{cases}
\label{eq:start_time}
\end{equation}
The request delay is
\begin{equation}
\widehat D_i^{\rm e2e}=
\max_{v\in\operatorname{Sink}(G_i)}
\widehat T_i^{\rm finish}(v)-\alpha_i.
\label{eq:e2e_delay}
\end{equation}
A successful plan must satisfy its request-local CPU checks and
\begin{equation}
\widehat D_i^{\rm e2e}+\xi_i^q\le D_i^{\max},
\label{eq:qos_commit}
\end{equation}
where $\xi_i^q$ is an optional queue-estimation margin. In the reported implementation, checked commitment covers request-local CPU feasibility, configured path reachability and latency, and delay-budget validation with queueing delay. Link state can enter path construction, but the paper does not rely on a separate universal residual-bandwidth or queue-capacity guarantee. The reported metrics retain QoS, delay, CPU balance, peak load, and per-request decision time separately so that the operating-point tradeoff remains visible.

\section{RQ-SAFE Online Orchestration}
\label{sec:framework}

\begin{figure*}[!t]
  \centering
  \includegraphics[width=0.95\textwidth]{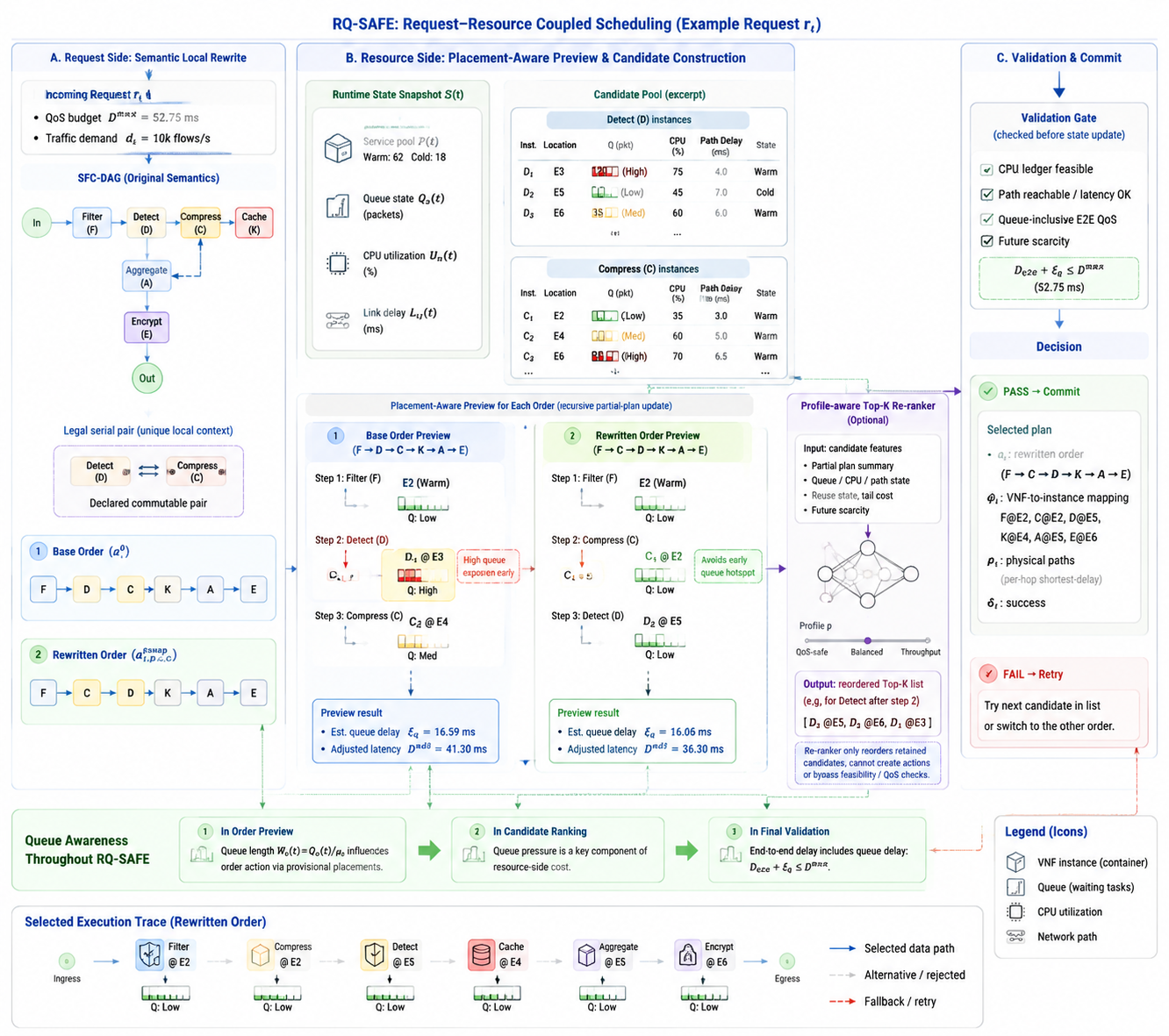}
  \caption{Request-level coupling mechanism of RQ-SAFE. A semantic local rewrite is evaluated by placement-aware preview under the current queue, resource, and path state. The retained order guides recursive candidate construction; profile-aware re-ranking refines close top-$K$ candidates; and the final plan is committed after CPU, path, and queue-aware QoS validation.}
  \label{fig:four_layer_detail}
\end{figure*}

Fig.~\ref{fig:four_layer_detail} expands the request-level path inside the orchestrator. It separates the input SFC-DAG, the local reordering choice, the shared queue/resource snapshot, the retained top-$K$ candidate lists, and the final validation gate. For each request, RQ-SAFE narrows the decision space through lightweight access control, semantic action extraction, placement-aware preview, recursive candidate construction, optional learning-assisted re-ranking, and checked state update. Algorithm~\ref{alg:four_layer_orchestration} summarizes the online path.

\begin{algorithm}[!t]
\caption{RQ-SAFE Online Orchestration}
\label{alg:four_layer_orchestration}
\footnotesize
\begin{algorithmic}[1]
\REQUIRE Request $r_i$, runtime state $\mathcal S(t)$, deployment profile $p$
\ENSURE Decision $\delta_i$ and selected plan $\mathcal P_i$
\STATE Apply lightweight risk screening
\STATE Extract $\mathcal A_i^{\rm ord}$ from semantic and structural rules
\FOR{each $a\in\mathcal A_i^{\rm ord}$}
    \STATE Build provisional plan $\widehat{\mathcal P}_i(a)$ and preview $\mathbf g_i(a,t)$
\ENDFOR
\STATE Retain the preferred legal action(s)
\FOR{each retained action $a$ in preview order}
    \FOR{each VNF $v$ in $\pi_i^a$}
        \STATE Form $\widetilde{\mathcal A}_{i,v}^{a}(t)$ and compute coupled scores
        \STATE Retain $\mathcal T^K_{i,v}$ and optionally re-rank it
        \STATE Select a candidate and update the request-local partial plan
    \ENDFOR
    \STATE Recompute end-to-end delay including queueing
    \IF{the complete request-local plan satisfies the success checks}
        \STATE Update global runtime state and \textbf{return} $\delta_i=1,\mathcal P_i$
    \ENDIF
\ENDFOR
\STATE \textbf{return} $\delta_i=0$
\end{algorithmic}
\end{algorithm}

\subsection{Executing the Coupled Model}

Section~\ref{sec:system_model} defines the semantic action set, placement-aware preview, recursive candidate preference, and queue-aware completion check. This section specifies how those objects are executed online rather than defining them again. Access control is used only as a lightweight prefilter: clearly unsuitable requests are rejected, whereas \emph{accept} and \emph{trial} requests enter the same coupled action/placement loop.

For each remaining request, the implementation extracts the legal actions in $\mathcal A_i^{\rm ord}$, builds a request-local provisional plan for each action, and retains at most $K_\pi$ actions according to their preview outcomes. Detailed placement then follows the retained order. At every VNF stage, the implementation evaluates bounded base, critical-path, tail/future, and resource/reuse groups corresponding to Eq.~\eqref{eq:coupled_cost}. The request-local partial plan is updated immediately after a tentative assignment. Consequently, predecessor paths, queue pressure, CPU pressure, reuse opportunities, and future scarcity are recomputed before the next VNF is ranked. This execution realizes the action-level and stage-level loops defined in Section~\ref{sec:system_model}.

\subsection{Learning-Assisted Re-Ranking}

The learning module is used only when several retained candidates are close under the deterministic score. It observes request and DAG context, the current partial plan, queue and resource pressure, QoS-related features, and candidate-level score components. Its output is a revised order within $\mathcal T^K_{i,v}$. Let $\sigma^0_{i,v}$ be the deterministic order and $\sigma^L_{i,v}$ the learned suggestion:
\begin{equation}
\sigma_{i,v}=\begin{cases}
\sigma^L_{i,v}, & \text{if the learned suggestion is accepted},\\
\sigma^0_{i,v}, & \text{otherwise}.
\end{cases}
\label{eq:learned_order}
\end{equation}
The module operates inside the retained candidate set and leaves semantic order actions to the explicit request-side mechanism. Its role is to adjust the tradeoff among close instance candidates before the common completion and QoS path.

\subsection{Success Checks, Fallback, and State Update}

Candidate construction uses request-local node and queue state. A candidate that lacks service availability or available CPU capacity is skipped, and the next retained candidate can be tried. After a complete plan is formed, end-to-end delay including queueing is recomputed. In the serial replay used for evaluation, global CPU and queue state are updated only after a successful request result. Failed requests leave the shared runtime state unchanged.

This structure gives learning a bounded and auditable influence. It may change which retained candidate is tried, but every selected plan reaches the same request-local CPU and queue-aware QoS path. Path reachability and link delay are incorporated during candidate construction. The commitment boundary is therefore explicit in the online path and is further detailed in the appendix.

\subsection{Per-Request Traceability}

For each request, the system records the exposed order actions, preview outcomes, retained candidates, actual candidate order, optional learned suggestion, fallback events, and final result:
\begin{equation}
\mathcal H_i=\left\langle
\mathcal A_i^{\rm ord},\{\mathbf g_i(a,t)\},a_i,
\{\mathcal T^K_{i,v},\sigma_{i,v}\}_{v\in\mathcal V_i},
\mathcal O_i^{\rm run},\delta_i
\right\rangle.
\label{eq:trace}
\end{equation}
The trace links a request-side action to its resource evidence and eventual outcome. It supports mechanism analysis and auditability, and matched experiments provide the performance evidence.

\section{Experimental Evaluation}

\subsection{Experimental Setup}
\label{subsec:experimental_setup}

We evaluate RQ-SAFE in an online edge SFC-DAG simulator with heterogeneous node capacities, link delays, warm/cold service pools, active tasks, and instance queues. Requests contain a dependency graph, traffic demand, explicit delay budget, priority, and request family. The workload suite covers public benchmark traces and high-concurrency or hybrid SFC-DAG snapshots with branches, merges, and local-order opportunities.

We compare with Greedy-CPU, Latency-Aware, DRL-DDQN, and GNN-DAG-Score. All methods use a common request parser, runtime-state builder, workload sequence, topology snapshot, service-pool state, and metric exporter. Each baseline retains its characteristic candidate rule or learned model. GNN-DAG-Score receives graph and runtime-state features from the shared environment; RQ-SAFE additionally receives the semantic action set produced by its request abstraction. The appendix reports the method input/action boundary, figure-specific settings, archive fingerprints, and run accounting.

The public-mixed families used in the repeated-run and factorial studies are denoted by W1--W4 when space is limited in tables and figures: W1 is Concurrent-balanced, W2 is Balanced-mixed, W3 is Queue-focused, and W4 is Safety-mixed. Each family is evaluated under matched seeds so that paired differences compare the same topology, request order, and service-pool snapshot. The primary metric is the QoS-compliant service ratio over all arrivals. We also report end-to-end delay, queue delay, peak CPU, CPU imbalance, hot-node ratio, and per-request decision time. Lower values are better for all metrics except the service ratio. The evaluation answers three research questions (RQs): \emph{RQ1}, how does RQ-SAFE perform on service outcome and delay; \emph{RQ2}, how does it change resource balance and per-request decision time; and \emph{RQ3}, how does request-side order flexibility interact with queue awareness in a direct factorial design?

\subsection{Research Question 1: Service Outcome Across Workloads}
\label{subsec:model_performance}

\begin{figure}[t]
    \centering
    \includegraphics[width=\columnwidth]{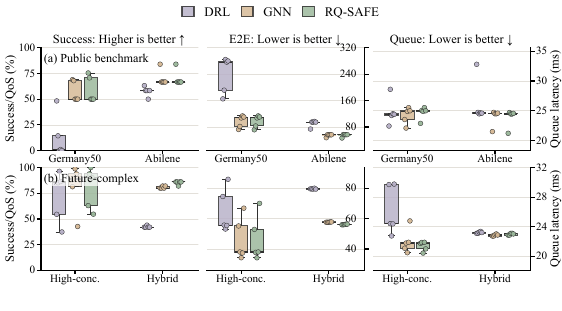}
    \caption{Cross-workload performance comparison. Panel (a) reports public benchmark workloads, and panel (b) reports future-complex workloads. Markers denote matched workload--seed configurations. QoS-compliant service ratio is higher is better; E2E and queue delay are lower is better.}
    \label{fig:cross_workload_tradeoff}
\end{figure}

Fig.~\ref{fig:cross_workload_tradeoff} positions DRL-DDQN, GNN-DAG-Score, and RQ-SAFE by service outcome and delay. Each point corresponds to a matched workload--seed run rather than an independent topology design; the plot therefore visualizes both average behavior and seed-level dispersion. On both public and future-complex groups, RQ-SAFE remains close to the graph-aware baseline and avoids the larger dispersion shown by DRL-DDQN in several groups. The explicit joint-scheduling pipeline therefore provides competitive service performance together with an inspectable request-action and candidate path.

Table~\ref{tab:rq1_paired_ci} adds paired accounting on the matched public-mixed workload--seed groups used in the multi-seed comparison. Values are RQ-SAFE minus GNN-DAG-Score; rate and budget metrics are reported in percentage points. The service-ratio interval crosses zero, and the end-to-end and queue-delay shifts are small in absolute value. RQ1 therefore shows that RQ-SAFE preserves the service-performance level of the graph-aware baseline while adding an explicit request--resource coupling path.

\begin{table}[t]
\centering
\caption{Paired Comparison with GNN-DAG-Score}
\label{tab:rq1_paired_ci}
\scriptsize
\setlength{\tabcolsep}{3pt}
\renewcommand{\arraystretch}{0.95}
\begin{tabular}{lrr}
\hline
Metric & Mean difference & 95\% paired CI \\
\hline
QoS-compliant ratio & $-1.01$ pp & $[-2.14,\,0.19]$ \\
Success ratio & $-1.01$ pp & $[-2.14,\,0.17]$ \\
End-to-end delay & $+0.21$ ms & $[0.04,\,0.36]$ \\
Queue delay & $+0.12$ ms & $[-0.07,\,0.31]$ \\
QoS-budget use & $+0.44$ pp & $[0.13,\,0.72]$ \\
\hline
\end{tabular}
\end{table}

The stricter topology-scale replay in the appendix reinforces this conclusion: RQ-SAFE and GNN-DAG remain in the leading group under strict request accounting. The resource-distribution results below distinguish their operating points.

\subsection{Research Question 2: Resource Balance and Decision Time}
\label{subsec:stability_repeated_runs}

\begin{figure}[t]
    \centering
    \includegraphics[width=\columnwidth]{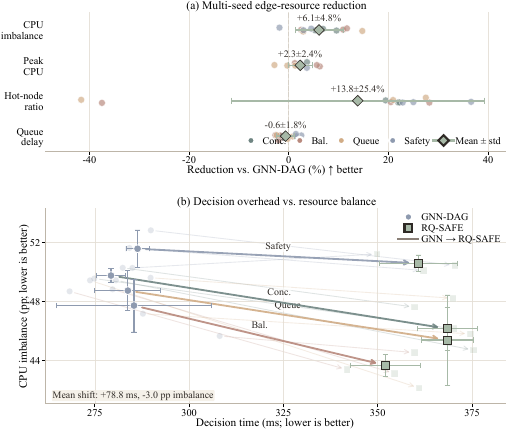}
    \caption{Multi-seed tradeoff between RQ-SAFE and GNN-DAG-Score on matched public-mixed SFC-DAG workloads. In (a), positive values mean that RQ-SAFE reduces a lower is better metric. In (b), moving right indicates higher per-request decision time and moving down indicates lower CPU imbalance.}
    \label{fig:repeated_run_stability}
\end{figure}

Fig.~\ref{fig:repeated_run_stability} uses four public-mixed workload families and three matched seeds per family. Relative to GNN-DAG-Score, RQ-SAFE achieves relative reductions of $6.1\%\pm4.8\%$ in CPU imbalance and $2.3\%\pm2.4\%$ in peak CPU. The hot-node ratio has a larger mean reduction, $13.8\%$, together with high seed-level dispersion ($25.4\%$ standard deviation), and is treated as secondary evidence. Queue delay is nearly neutral in this comparison ($-0.6\%\pm1.8\%$).

Across the matched workload--seed pairs, the mean shift is $+78.8$ ms in per-request decision time and $-3.0$ percentage points in CPU imbalance. Paired tests show reductions of $2.995$ percentage points in CPU imbalance (95\% CI $[-4.310,-1.749]$, Wilcoxon $p=0.0010$) and $2.193$ percentage points in peak CPU (95\% CI $[-3.452,-0.887]$, $p=0.0146$). Per-request decision time increases by $78.772$ ms (95\% CI $[71.847,85.128]$, $p<0.001$). The queue-delay difference is not significant ($0.125$ ms, 95\% CI $[-0.073,0.305]$, $p=0.2036$). RQ-SAFE thus incurs higher control-plane per-request decision time to reach a less concentrated CPU operating point while preserving a comparable service outcome. This overhead should be interpreted as orchestration time rather than data-plane packet latency; the supplementary decomposition below shows that it is dominated by checked placement-and-validation search.

Profile-stress and queue-estimation-error results are reported in the appendix because they characterize deployment tuning rather than the central cross-method result.

\subsection{Research Question 3: Request--Queue Coupling}
\label{subsec:factorial_coupling}

\begin{figure}[t]
    \centering
    \includegraphics[width=\columnwidth]{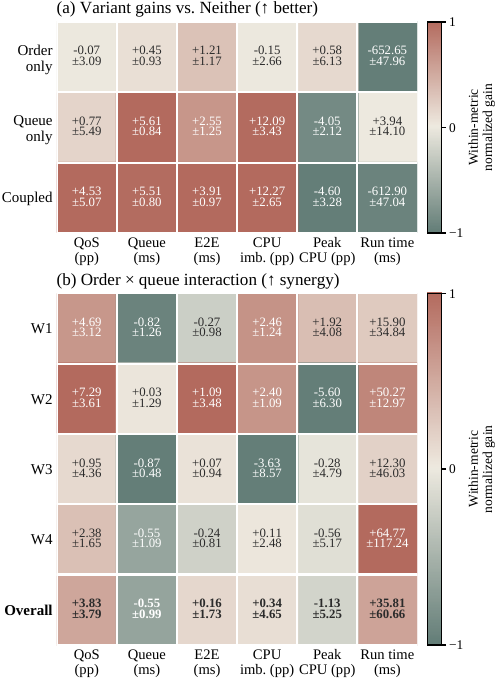}
    \caption{Direct $2\times2$ factorial evaluation of request-side order flexibility and queue-aware decision making. (a) Gains of Order only, Queue only, and Coupled relative to Neither; signs are converted so that positive is preferable for every metric. (b) Order--queue interaction $I_m=s_m(C-O-Q+N)$, where $s_m=1$ for higher is better metrics and $s_m=-1$ otherwise. Entries show mean$\pm$standard deviation. W1--W4 each contain three matched seeds; Overall contains all 12 workload--seed groups. Colors are normalized within each metric, while annotations retain raw percentage-point or millisecond values.}
    \label{fig:factorial_coupling}
\end{figure}

Fig.~\ref{fig:factorial_coupling} directly isolates the two factors underlying the joint-scheduling claim. This experiment is a mechanism isolation test rather than an external baseline ranking. The four matched variants are Neither, Order only, Queue only, and Coupled, giving 48 completed runs over four workload families and three seeds. Table~\ref{tab:action_coverage_adoption} summarizes the request-level coverage of the local-order mechanism in the factorial suite; adoption is measured only for order-enabled variants. In the table, Previewed counts requests for which the default order and a legal alternative were both evaluated, Better alt. counts cases where the non-default order produced a better provisional outcome, and Applied counts cases where that alternative became the selected order. In the order-enabled conditions, all 1608 eligible requests entered the serial placement preview, and 557 selected and applied the better legal alternative. The coupled variant applies local rewrites more often than the order-only variant, showing that queue/resource state changes how request-side freedom is used.

\begin{table}[t]
\centering
\caption{Local-Order Action Coverage and Adoption}
\label{tab:action_coverage_adoption}
\scriptsize
\setlength{\tabcolsep}{2.6pt}
\renewcommand{\arraystretch}{0.95}
\begin{tabular}{lrrrr}
\hline
Variant & Requests & Previewed & Better alt. & Applied \\
\hline
Order only & 804 & 804 & 221 (27.5\%) & 221 (27.5\%) \\
Coupled & 804 & 804 & 336 (41.8\%) & 336 (41.8\%) \\
Order-enabled total & 1608 & 1608 & 557 (34.6\%) & 557 (34.6\%) \\
\hline
\end{tabular}
\end{table}

Relative to Neither, Coupled improves the QoS-compliant service ratio by $4.53\pm5.07$ percentage points. It also reduces queue delay by $5.51$ ms, end-to-end delay by $3.91$ ms, and CPU imbalance by $12.27$ percentage points. These gains are accompanied by a $4.60$-percentage-point increase in peak CPU and about $612.90$ ms of additional per-request decision time. The coupled condition serves more requests within QoS and spreads the admitted work more evenly; the larger served load can raise the absolute peak, while the extra previews add computation.

For metric $m$, the signed interaction is
\begin{equation}
I_m=s_m(C_m-O_m-Q_m+N_m),
\label{eq:factorial_interaction}
\end{equation}
where $N$, $O$, $Q$, and $C$ denote the four variants and $s_m$ makes positive values preferable. The interaction term measures the extra gain that appears when the two factors are enabled together, beyond the sum of their separate effects. The overall QoS interaction is $3.83\pm3.79$ percentage points, with a 95\% bootstrap confidence interval of $[1.77,5.89]$ and a paired sign-flip $p$-value of $0.0059$. Table~\ref{tab:workload_interaction} shows that the QoS interaction is positive in all four workload families. The W1--W4 labels follow the workload definitions in the setup, and each workload contains three matched seeds. Values are signed gains; for delay and CPU imbalance, positive values mean reductions. The additional QoS gain appears when legal order flexibility and current queue state are active together.

\begin{table}[t]
\centering
\caption{Workload-Level Order--Queue Interaction}
\label{tab:workload_interaction}
\scriptsize
\setlength{\tabcolsep}{2.8pt}
\renewcommand{\arraystretch}{0.95}
\begin{tabular}{lrrr}
\hline
Workload & QoS & Queue & CPU imb. \\
 & (pp) & (ms) & (pp) \\
\hline
W1 Concurrent-balanced & $+4.69$ & $-0.82$ & $+2.46$ \\
W2 Balanced-mixed & $+7.29$ & $+0.03$ & $+2.40$ \\
W3 Queue-focused & $+0.95$ & $-0.87$ & $-3.63$ \\
W4 Safety-mixed & $+2.38$ & $-0.55$ & $+0.11$ \\
Overall & $+3.83$ & $-0.55$ & $+0.34$ \\
\hline
\end{tabular}
\end{table}

The factorial design maps directly to the model. Order only activates $\mathcal A_i^{\rm ord}$ and the action-level preview while masking queue information from the decision. Queue only activates the queue terms and queue-aware decision path while retaining the base order. Coupled closes both the action-level loop in Eq.~\eqref{eq:order_preview} and the stage-level loop in Eq.~\eqref{eq:recursive_joint_placement}. The positive QoS interaction is therefore evidence for the joint mechanism rather than a generic benefit from adding two independent modules.

The direct delay and CPU-imbalance reductions arise mainly from the queue-aware main effect; their interaction intervals include zero. The evidence is consequently metric-specific: order--queue coupling contributes the clearest non-additive gain to QoS-compliant service outcome, while queue awareness supplies most of the immediate delay and balancing benefit. Additional module-removal and request-trace results are reported in the appendix.

\subsection{Supplementary Order-Budget and Runtime Checks}
\label{subsec:stage3_supplement}

We add two lightweight checks to clarify the operating range of the mechanism. First, we vary the retained order-action budget $K_\pi$. In the action-rich suite, each eligible request exposes the default order plus one retained non-default local rewrite. Thus $K_\pi=1$ keeps only the default order, whereas $K_\pi\ge2$ saturates the retained order-action set. Moving from $K_\pi=1$ to $K_\pi=2$ enables placement-aware preview for 804 eligible coupled-variant requests and leads to 336 selected non-default local orders. Larger budgets ($K_\pi=3,5$) leave the service and action-adoption outcomes unchanged in this suite. This result shows order-action budget saturation for the evaluated action-rich workload.

Second, we decompose the measured per-request decision time using the recorded timing fields. The dominant component is placement-and-validation search, which accounts for $256.01$ ms on average, or $67.84\%$ of the mean decision time. Semantic order abstraction and access-control screening are small in comparison, at $0.07$ ms and $1.22$ ms, respectively. The higher decision time observed in the cross-method comparison is therefore mainly the cost of checked candidate construction and validation, not of extracting the bounded local-order action set. Full tables are reported in the appendix.

\begin{figure*}[!t]
  \centering
  \includegraphics[width=0.88\textwidth]{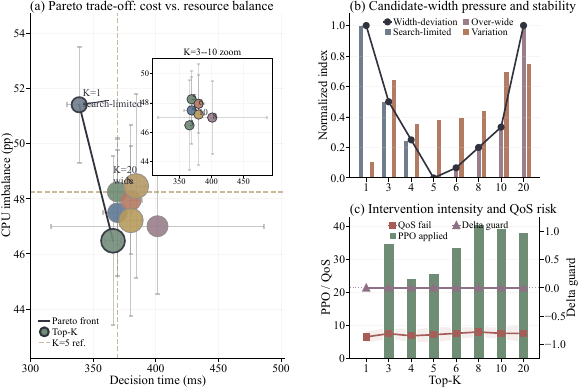}
  \caption{Top-$K$ candidate-width sensitivity. Panel (a) shows the per-request decision-time and CPU-imbalance tradeoff; panel (b) summarizes candidate-width pressure and stability; panel (c) reports intervention intensity and QoS-risk traces.}
  \label{fig:app_topk_sensitivity}
\end{figure*}

Fig.~\ref{fig:app_topk_sensitivity} reports the retained candidate-width sensitivity. A moderate retained width provides enough search coverage for stable candidate selection, while a wider list increases control-plane computation. This supports the default top-$K$ setting used in the main experiments.

\begin{figure*}[!t]
  \centering
  \includegraphics[width=0.88\textwidth]{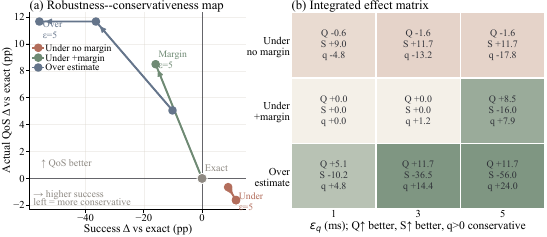}
  \caption{Queue-estimation-error robustness. Panel (a) maps QoS recovery against success-rate cost relative to exact queue estimation; panel (b) integrates QoS change, success change, and validation-queue shift for $\epsilon_q=1,3,5$ ms.}
  \label{fig:app_queue_error_robustness}
\end{figure*}

Fig.~\ref{fig:app_queue_error_robustness} tests the effect of bounded queue-estimation error. Under-estimating queueing delay can reduce actual QoS compliance, while adding a validation margin recovers QoS at the cost of a more conservative success rate. This result clarifies the runtime boundary of the queue-aware validation used by RQ-SAFE.

\subsection{Interpreting the Joint Scheduling Mechanism}
\label{subsec:joint_interpretation}

The model joins the request and resource sides in a stronger sense than placing two objective terms in one weighted sum. The request-side action changes the order in which predecessor context, paths, service instances, and remaining QoS slack are exposed. The resource-side state then determines the provisional plan generated under that action. A different action can therefore change both the candidate set in Eq.~\eqref{eq:filtered_candidates} and the value of a candidate in Eqs.~\eqref{eq:req_cost}--\eqref{eq:res_cost}. Conversely, every tentative resource choice updates the partial plan and changes the information used for the next request stage. The action and placement variables are connected through state transitions, not only through a final scalar score.

This distinction explains why the action-level and stage-level loops are both needed. The action-level loop prevents a semantically legal rewrite from being selected only because it looks attractive in the request graph. It must first produce a feasible and competitive provisional placement under the current service pools and queues. The stage-level loop prevents the retained action from becoming a fixed template once detailed placement begins. CPU pressure, waiting time, path context, reuse opportunities, and future scarcity are updated after each tentative assignment. Thus, request-side freedom determines what resource alternatives become reachable, while resource-side feedback determines how that freedom is used.

The profile weights in Eq.~\eqref{eq:coupled_cost} change the operating point without changing this coupling structure. A conservative profile can place more emphasis on completion slack, critical stages, and tail exposure. A throughput-oriented profile can tolerate more projected pressure when reusable service capacity is available. Learning-assisted re-ranking operates one level lower: it adjusts the order of close retained candidates after the explicit score has exposed the relevant tradeoff. It refines instance ordering while the semantic action set, provisional placement, and recursive state update remain explicit. The model can therefore support multiobjective adjustment while preserving a readable path from the incoming request to the final resource outcome.

This interpretation also guides the experiments. Cross-method comparisons evaluate the service and resource operating point of the full process. The factorial design tests whether order flexibility becomes more valuable when queue information is active. Request-level traces then show which action, candidate order, and fallback path produced each observed outcome. Together, these views test the same mechanism at system, factor, and decision levels.

\subsection{Traceability of the Coupled Decision Path}
\label{subsec:traceability}

\begin{figure}[t]
    \centering
    \includegraphics[width=\columnwidth]{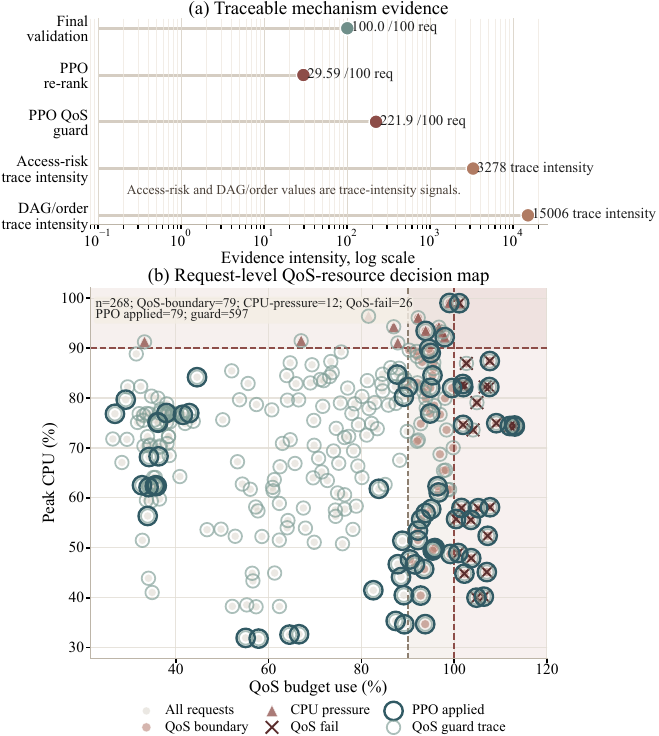}
    \caption{Traceability of the recorded decision path. (a) Two evidence groups are displayed on a logarithmic axis. Final validation, learned re-ranking, and QoS-guard events are request-normalized rates per 100 requests; access-risk and DAG/order entries are raw trace-record intensities and are not magnitude-comparable with those rates. (b) Request-level QoS-budget use and peak CPU. Circled points contain an accepted learned re-ranking record; the circle indicates mechanism involvement rather than a causal performance gain.}
    \label{fig:traceability}
\end{figure}

Panel~(a) reports evidence coverage rather than a ranking of mechanism frequency. The request-normalized group shows that the final result path is recorded for every request and that accepted learned re-ranking appears at $29.59$ events per 100 requests. QoS-guard events are more frequent because one request can generate several candidate-level checks. The access-risk and DAG/order entries belong to a second group: they count intermediate trace records produced while constructing actions and partial plans. Their larger values therefore indicate finer logging granularity rather than more requests or greater algorithmic importance.

Panel~(b) locates 268 final request outcomes in the QoS--resource operating space, including 79 requests near the QoS boundary, 12 under high CPU pressure, and 26 QoS failures. For a circled request, the record links the selected local action, its placement preview, the deterministic and learned candidate orders, fallback events, and the final outcome. A circle indicates that learning participated in the recorded candidate path. The paired comparisons and factorial experiment provide the performance evidence; Fig.~\ref{fig:traceability} establishes that the intermediate decision path remains observable.

A decoded request trace is reported in the appendix. In that record, a legal local rewrite is exposed, evaluated through a queue-aware provisional placement, used to condition candidate scoring, and accepted only after the queue-aware end-to-end budget is checked. The main role of Fig.~\ref{fig:traceability} is therefore to show that this audit path exists for real requests, while the performance conclusions come from the matched comparisons above.

\section{Conclusion and Future Work}
\label{sec:conclusion}

This paper studied online orchestration for partially ordered edge SFC-DAGs. RQ-SAFE turns semantic partial-order freedom into bounded local actions, evaluates each action through placement-aware preview, and continues detailed placement under the retained order. The resulting two-level loop connects request-side completion and QoS requirements with queues, paths, service pools, projected CPU pressure, and future candidate availability.

The experiments show that this joint-scheduling design maintains a service-performance level comparable to graph-aware baselines and moves deployments toward lower CPU concentration with higher control-plane decision time. The direct factorial experiment identifies a positive order--queue interaction in QoS-compliant service outcome, showing that queue state changes how legal order freedom is used. Supplementary checks show that the retained order-action set saturates once the available non-default local rewrite is exposed in the action-rich suite, and that the measured decision cost is dominated by checked placement-and-validation search. Learning-assisted re-ranking remains a bounded candidate refinement, while request actions, partial plans, fallbacks, and outcomes remain traceable.

Future work will extend queue estimation with production telemetry, calibrate marginal reuse cost, and evaluate larger physical and containerized testbeds. Migration, autoscaling, and lifecycle-aware activation can be placed around the same bounded action, recursive placement, and trace interfaces.



\end{document}


\appendices
\setcounter{table}{0}
\renewcommand{\thetable}{A.\Roman{table}}

\section{Commitment Checks and Queue-Aware QoS Analysis}
\label{app:feasibility_qos}

This appendix states the commitment properties used by the main text and the exact runtime boundary supported by the released implementation. Candidate construction operates on request-local resource and queue state. A retained instance can be selected only when its service is available and its projected CPU increment fits the request-local CPU ledger. After a complete plan is formed, queue-aware end-to-end delay is recomputed. In the serial replay used for evaluation, global CPU and queue state advance only after a successful request result. Link reachability and latency enter candidate/path construction; the reported commitment boundary covers request-local CPU feasibility, path reachability/latency construction, and queue-aware QoS validation rather than a separate universal residual-bandwidth or queue-capacity guarantee.

\subsection{CPU Feasibility Under Atomic Commitment}

\noindent\textbf{Proposition A.1. CPU feasibility under atomic commitment.}
Assume that resource updates are performed serially or through atomic commitment. If a request is committed by the validation layer, then the committed CPU usage of every physical node remains within its CPU capacity.

\noindent\textit{Proof sketch.}
For each VNF-to-instance mapping \((v,s)\) in a candidate deployment plan, let \(n(s)\) denote the physical node that hosts instance \(s\). During validation, the orchestrator maintains a temporary resource ledger \(\widetilde{U}^{\mathrm{cpu}}_n(t)\). A candidate can be selected only if
\[
\widetilde{U}^{\mathrm{cpu}}_{n(s)}(t)+\Delta r^{\mathrm{cpu}}_{v,s}
\le C^{\mathrm{cpu}}_{n(s)}-\xi^{\mathrm{cpu}}_p ,
\]
where \(\Delta r^{\mathrm{cpu}}_{v,s}\) is the effective CPU increment caused by assigning VNF \(v\) to instance \(s\), and \(\xi^{\mathrm{cpu}}_p\ge 0\) is the profile-dependent safety margin. Candidates outside this bound are skipped. When no feasible candidate exists for any VNF, the request is rejected and the physical resource ledger remains unchanged. Hence every committed mapping satisfies
\[
\widetilde{U}^{\mathrm{cpu}}_{n(s)}(t)+\Delta r^{\mathrm{cpu}}_{v,s}
\le C^{\mathrm{cpu}}_{n(s)} .
\]
Under serial or atomic commitment, atomicity prevents intermediate conflicting updates from invalidating the checked condition. The committed CPU usage of each node therefore remains within its capacity. \proofend

\subsection{Queue-Aware QoS Compliance Under Exact Queue Estimation}

\noindent\textbf{Proposition A.2. Queue-aware QoS compliance under exact queue estimation.}
If the queueing delay used in validation is exact, then any request committed by the validation layer satisfies its end-to-end delay budget.

\noindent\textit{Proof sketch.}
For request \(r_i\), the validation layer recomputes the queue-aware end-to-end delay of the CPU-feasible plan:
\[
\widehat{D}^{\mathrm{e2e}}_i
= D^{\mathrm{e2e}}_i(\mathcal{P}^{\mathrm{safe}}_i,\mathcal{S}(t)).
\]
The final commitment condition is
\[
\widehat{D}^{\mathrm{e2e}}_i \le D^{\max}_i .
\]
If the queueing-delay estimates are exact, then \(\widehat{D}^{\mathrm{e2e}}_i\) equals the actual queue-aware end-to-end delay after deployment. Every committed request therefore satisfies the delay budget \(D^{\max}_i\). \proofend

\subsection{Robust QoS Compliance Under Bounded Queue-Estimation Error}

\noindent\textbf{Proposition A.3. Robust QoS compliance with queue-margin validation.}
Suppose that the queueing-delay estimation error of each selected VNF instance is bounded by
\[
\left|D^{\mathrm{queue}}_{i,v}-\widehat{D}^{\mathrm{queue}}_{i,v}\right|
\le \epsilon_q,\quad \forall v\in\mathcal{V}_i .
\]
If the validation layer commits a request only when
\[
\widehat{D}^{\mathrm{e2e}}_i + |\mathcal{V}_i|\epsilon_q \le D^{\max}_i ,
\]
then the actual end-to-end delay of the committed request also satisfies the QoS budget.

\noindent\textit{Proof sketch.}
For any source-to-sink path in the SFC-DAG, the total queueing-delay estimation error is upper-bounded by the number of VNFs on the path multiplied by \(\epsilon_q\). Since any path contains at most \(|\mathcal{V}_i|\) VNFs, the queue-aware end-to-end delay satisfies
\[
D^{\mathrm{e2e}}_i
\le \widehat{D}^{\mathrm{e2e}}_i + |\mathcal{V}_i|\epsilon_q .
\]
If the validation layer reserves the margin \(|\mathcal{V}_i|\epsilon_q\), then
\[
D^{\mathrm{e2e}}_i
\le \widehat{D}^{\mathrm{e2e}}_i + |\mathcal{V}_i|\epsilon_q
\le D^{\max}_i .
\]
Thus the deployed request remains QoS-compliant under bounded queue-estimation error. \proofend

\section{Semantic Local-Order Actions and Placement-Aware Preview}
\label{app:order_preview}

The request-side mechanism exposes only local rewrites that are supported by both declared VNF semantics and a conservative structural rule. It exposes a compact action set instead of enumerating all topological orders. Let $\mathcal C_i$ denote the admitted local pairs. The action set is
\[
\mathcal A_i^{\rm ord}=\{a_i^0\}\cup
\{a_{i,u,v}^{\rm swap}:(u,v)\in\mathcal C_i\}.
\]

The implementation uses two rule families. For a linear request, an eligible pair must be internal and adjacent in the base orchestration order. For a branch or general DAG, an eligible pair must be a directed serial pair on one branch. Its local predecessor and successor context must be unique, and merge and tail nodes are excluded. Independent siblings may be recorded as structural information, but they are not turned into semantic swap actions. Unknown type pairs default to non-commutable.

For a branch-local serial segment $p\!\rightarrow\!u\!\rightarrow\!v\!\rightarrow\!s$, a valid action rewrites only that segment as $p\!\rightarrow\!v\!\rightarrow\!u\!\rightarrow\!s$. All other edges remain unchanged. The rewrite is admitted only when the configured type-pair rule and the local structural guard both pass. The rule families used by the released implementation are summarized in Table~\ref{tab:implemented_order_rules}.

\begin{table*}[!tbp]
\centering
\caption{Implemented local-order rule families. Type names follow the benchmark service taxonomy.}
\label{tab:implemented_order_rules}
\scriptsize
\setlength{\tabcolsep}{3pt}
\renewcommand{\arraystretch}{1.02}
\begin{tabular}{p{0.17\linewidth}p{0.27\linewidth}p{0.25\linewidth}p{0.24\linewidth}}
\toprule
Rule family & Structural condition & Enabled type pairs & Conservative exclusions \\
\midrule
Linear local rewrite & Internal adjacent pair in the base order; surrounding predecessor/successor context remains fixed & cache--forward, cache--aggregate, aggregate--forward & Source/sink endpoints, undeclared pairs, and state/side-effect conflicts \\
Branch-local serial rewrite & Directed pair on one branch with unique local predecessor and successor context & detect--cache, detect--compress, cache--compress & Merge or tail nodes, independent siblings, ambiguous branch context, and undeclared pairs \\
\bottomrule
\end{tabular}
\end{table*}

Each action is evaluated by the deterministic placement-aware preview used in the online request path. The preview invokes the current queue/resource-sensitive candidate process and records whether a provisional plan can be formed, together with processing/activation delay, queue pressure, network delay, adjusted end-to-end delay, and QoS slack. The original order is retained unless a legal rewrite produces a preferable provisional outcome. Request-side order selection is therefore explicit and deterministic. The learned module acts later, only on retained instance candidates.

\section{Queue Estimation Error Analysis}
\label{app:queue_error}

The main paper uses a lightweight queueing-delay estimator based on the observable queue length and service rate of each VNF instance:
\[
\widehat{W}_s(t)=\frac{Q_s(t)}{\mu_s}.
\]
This estimator is suitable for online orchestration because it uses runtime information that can be collected from the service pool. It can still deviate from the actual waiting time when service times are heterogeneous, requests arrive in bursts, background tasks are present, or instance-level concurrency changes quickly.

For VNF \(v\) assigned to instance \(s\), we write the queueing-delay error as
\[
D^{\mathrm{queue}}_{i,v}=\widehat{D}^{\mathrm{queue}}_{i,v}+e_{i,v},
\]
where \(e_{i,v}\) is the estimation error. A conservative bounded-error assumption is
\[
|e_{i,v}| \le \epsilon_q .
\]
Under this assumption, the total error of a chain-like request is at most \(|\mathcal{V}_i|\epsilon_q\). For a general SFC-DAG, the end-to-end delay is determined by the maximum source-to-sink completion time. The bound can therefore be tightened by using the maximum path length:
\[
D^{\mathrm{e2e}}_i
\le \widehat{D}^{\mathrm{e2e}}_i + L_i^{\max}\epsilon_q,
\]
where \(L_i^{\max}\) is the maximum number of VNFs on any source-to-sink path of \(G_i\). Since \(L_i^{\max}\le |\mathcal{V}_i|\), the bound in Proposition A.3 is conservative.

In practice, \(\epsilon_q\) can be estimated from validation traces or set as a profile-dependent margin. Conservative profiles may use a larger queue margin, while throughput-oriented profiles may use a smaller margin. This gives a direct tradeoff between QoS robustness and access-control aggressiveness.

\section{Bounded Decision Properties of Learning-Assisted Top-\texorpdfstring{$K$}{K} Re-Ranking}
\label{app:learning_reranking}

The main text treats this component as learning-assisted re-ranking. The released online path loads a frozen ranking artifact and applies it to a retained instance-candidate list. The paper's learning claim is scoped to this runtime role and to the retained-candidate neighborhood used by online orchestration.

\subsection{Bounded Online Decision Space and Inputs}
For VNF $v$ of request $r_i$, the learning module can select only an index in the retained list:
\[
\mathcal A^{\rm LR}_{i,v}=\{1,2,\ldots,|\mathcal T^K_{i,v}|\},
\qquad |\mathcal T^K_{i,v}|\le K.
\]
The action space therefore grows with the configured retained width, not with the total number of nodes or VNF instances. Online inputs contain only pre-commit information: request and DAG summaries, the current partial plan, queue/resource state, deterministic candidate scores, and candidate-level path, reuse, tail, and scarcity features. Realized post-deployment outcomes are not online inference features.

\subsection{Runtime Boundary and Fallback}
The learning component neither creates candidates nor selects request-side order actions. It returns a revised ordering of $\mathcal T^K_{i,v}$. If the suggestion is unavailable or rejected by the runtime control, the deterministic order is retained. Candidate feasibility, request-local CPU accounting, complete-plan latency computation, and the successful state-update path are shared with the rule-only mode.

\noindent\textbf{Proposition C.1. Learned re-ranking preserves the enabled success checks.}
If global state is updated only after a complete request-local plan passes service availability, CPU-capacity, and queue-aware QoS checks, then changing the order of candidates inside $\mathcal T^K_{i,v}$ preserves the same state-update boundary.

\noindent\textit{Proof sketch.}
The learned output is an ordering of an already retained set. Each candidate is processed by the same request-local construction path. A candidate that lacks service availability or available CPU capacity is skipped. After a complete plan is formed, queue-aware end-to-end delay is recomputed. If the plan fails, the next retained candidate/order can be tried or the request is rejected. Global state is advanced only for a successful result. Re-ranking therefore changes search order but not the enabled success conditions. \proofend

\subsection{Scope of the Learning Claim}

The results above establish the online boundary needed by this paper: learning can only reorder an already retained candidate set, and every resulting order enters the same request-local CPU, completion, fallback, and state-update path. The learning claim is limited to bounded retained-candidate re-ranking; global optimality, monotonic policy improvement, and calibrated approximation ratios are outside this runtime role. Candidate-width quality and overhead are evaluated directly by the sensitivity experiment and the complexity analysis.

\section{Computational Complexity}
\label{app:complexity}

This appendix states the online complexity with the candidate-order and fallback loops made explicit. Let \(|\mathcal{V}_i|\) and \(|\mathcal{E}_i|\) be the numbers of VNFs and virtual dependency edges in request \(r_i\). Let \(|\mathcal{N}|\) and \(|\mathcal{L}|\) be the numbers of physical nodes and links. Let \(M\) denote the average number of type-matched candidate instances per VNF before top-\(K\) truncation. Let \(K_\pi\) be the maximum number of retained candidate execution orders, \(K\) be the local top-\(K\) candidate width per VNF, and \(B\le K\) be the maximum number of candidate retries per VNF and per retained order during validation. The top-\(K\) list in this paper is a local per-VNF candidate list, so the online path builds retained local lists rather than enumerating all complete plans in \(K^{|\mathcal{V}_i|}\).

\subsection{Access Control and Order Abstraction}

The access-control layer computes queue, resource, latency, and structural risk scores. If candidate availability is checked over an average of \(M\) instances per VNF, the cost is
\[
O(|\mathcal{V}_i|M).
\]
The SFC-DAG/order abstraction layer extracts commutable segments, branch/merge labels, critical paths, and tail-sensitive VNFs. Let \(C_i\) be the number of candidate local-order actions before pruning. The implementation retains at most \(K_\pi\) orders, so the abstraction and pruning cost can be written as
\[
O(|\mathcal{V}_i|+|\mathcal{E}_i|+C_i\log K_\pi),
\]
or simply \(O(|\mathcal{V}_i|+|\mathcal{E}_i|+C_i)\) when \(K_\pi\) is treated as a small deployment constant.

\subsection{Candidate Generation Under Retained Orders}

For each retained order, the placement layer filters and scores candidate instances for every VNF. Filtering and scoring over the type-matched pool costs
\[
O(|\mathcal{V}_i|M),
\]
and retaining a local top-\(K\) list by heap or partial sorting costs
\[
O(|\mathcal{V}_i|M\log K).
\]
Across all retained orders, this part contributes
\[
O\!\left(K_\pi |\mathcal{V}_i|M\log K\right).
\]
This expression is the cost of building local candidate lists. It is not a full-plan enumeration cost.

\subsection{Path Search, Re-Ranking, and Validation Fallback}

Let
\[
P_{\rm sp}=O(|\mathcal{L}|+|\mathcal{N}|\log |\mathcal{N}|)
\]
be the cost of one configured shortest-path search over the current link-state snapshot. For a fixed retained order, path construction over the virtual edges costs \(O(|\mathcal{E}_i|P_{\rm sp})\). During candidate construction or commitment, a candidate may fail a service-availability, CPU-capacity, configured-path, or QoS check. The validator may then try the next retained candidate. With at most \(B\) retries per VNF and per retained order, the fallback validation cost is bounded by
\[
O\!\left(B|\mathcal{V}_i| + B|\mathcal{E}_i|P_{\rm sp}\right),
\]
where the first term covers candidate and ledger checks and the second term covers the worst case in which retries trigger additional path searches. The learned module reorders only the retained local lists, so with fixed neural hidden size its cost is
\[
O(K_\pi |\mathcal{V}_i|K).
\]
Combining retained orders, local candidate construction, re-ranking, fallback validation, and path search gives the per-request upper bound
\[
O\!\left(
K_\pi\left[
|\mathcal{V}_i|M\log K
+ |\mathcal{V}_i|K
+ B|\mathcal{V}_i|
+ B|\mathcal{E}_i|P_{\rm sp}
\right]
\right),
\]
plus the access-control and structural-abstraction cost above. When \(K_\pi\), \(K\), and \(B\) are bounded deployment parameters, the dominant terms are candidate filtering/scoring and, when enabled, path search. Increasing the retained order count, local candidate width, or retry budget improves search coverage but directly increases per-request decision time; this is the tradeoff characterized by the retained-candidate-width sensitivity and topology-scale experiments.

\section{Experimental Settings and Runtime Parameters}
\label{app:runtime_parameters}

This appendix summarizes the settings used in the reported experiments. The evaluation uses mixed edge SFC-DAG benchmark snapshots rather than one fixed toy topology. All methods within the same comparison use the same topology snapshot, workload trace, seed, service-pool state, and request count. Exact YAML paths, generated configuration files, request-payload fingerprints, runtime-trace fingerprints, model aliases, and run modes are stored in the corresponding figure archives.

The current experiments use CPU as the binding node resource in request-local feasibility checks. Memory is retained as an extensible resource dimension. Link state and latency are used during configured path construction; the reported runtime uses link reachability and latency during configured path construction rather than a separate universal residual-bandwidth hard guarantee. Peak node CPU denotes the maximum CPU utilization among physical edge nodes after a deployment attempt. The hot-node ratio is the fraction of nodes whose CPU utilization exceeds the hotspot threshold. The CPU-imbalance metric is the peak-to-minimum CPU spread over physical nodes. Shadow evaluation logs a learned suggestion in shadow mode, where the logged suggestion is not applied to the request path.

For the multi-seed edge-deployment comparison in Fig.~4, the workloads are public-mixed SFC-DAG request sets on the SNDlib Germany50 topology. The four workload families are Concurrent-balanced, Balanced-mixed, Queue-focused, and Safety-mixed. Each family is evaluated with matched seeds 831, 832, and 833 for both RQ-SAFE and GNN-DAG. The generated archive verifies that the request payload, runtime trace, and metric fingerprints differ across seeds. Main-text Fig.~5 uses a direct $2\times2$ order-flexibility-by-queue-awareness factorial suite with the same four workload families, three matched seeds, and four variants (Neither, Order only, Queue only, and Coupled), giving 48 completed runs and 12 complete matched groups. Fig.~6 uses the request-level traceability and mechanism-evidence suite. The exact request counts are figure-specific and are recorded in the released archives.

\begin{table*}[!tbp]
\caption{Reported Simulation Settings}
\label{tab:simulation_settings}
\centering
\scriptsize
\setlength{\tabcolsep}{3pt}
\renewcommand{\arraystretch}{0.94}
\begin{tabular}{p{0.24\textwidth}p{0.68\textwidth}}
\hline
\textbf{Parameter} & \textbf{Value} \\
\hline
Benchmark source & Mixed edge SFC-DAG benchmark snapshots. Fig.~4 uses public-mixed SFC-DAG workloads on the SNDlib Germany50 topology. Main-text Fig.~5 uses matched order--queue factorial workloads, and Fig.~6 uses traceability snapshots recorded in the figure archives. \\
Physical topology & Snapshot-specific directed edge topology; all methods in the same comparison use the same topology and link-state snapshot. \\
Node CPU capacity & Snapshot-specific normalized CPU capacity; CPU is the binding node resource in request-local success checks. \\
Memory capacity & Retained as an extensible resource dimension; reported hard node-feasibility checks use CPU. \\
VNF types & Configured service types include filtering, detection, encryption/decryption, caching, compression, forwarding, monitoring, policy checking, aggregation, re-ranking, and sanitization functions. \\
Warm VNF instances & Initialized from each benchmark service-pool snapshot; cold-start candidates are considered only when resources allow activation. \\
VNF CPU demand & Normalized per-type or per-instance demand from the selected benchmark configuration. \\
VNF processing delay & Per-type processing delay from the selected benchmark configuration. \\
Cold-start delay & Configuration-specific activation delay for cold candidates. \\
Link state and delay & Heterogeneous link-state and link-delay fields are read from the selected topology snapshot. Path reachability and latency enter candidate construction; link reachability and latency are used without a separate universal residual-bandwidth hard guarantee. \\
Traffic demand model & Edge-specific traffic variables \(\ell_{i,u,v}\) and \(b_{i,u,v}\) are supported; request-level defaults are used when branch-specific traffic is unavailable. \\
QoS delay budgets & Explicit QoS budgets are stored with each workload snapshot and are reused by all compared methods under the same workload seed. \\
Request families & Latency-sensitive, security-sensitive, long-chain, parallel, branch/merge, commutable-order, and tail-latency-sensitive services. \\
Fig.~4 workload families & Concurrent-balanced, Balanced-mixed, Queue-focused, and Safety-mixed public-mixed SFC-DAG workloads. \\
Fig.~4 seeds & Matched seeds 831, 832, and 833 for each workload family and for both RQ-SAFE and GNN-DAG. \\
Seed verification & The Fig.~4 archive records generated configuration fingerprints, request-payload fingerprints, runtime-trace fingerprints, and metric fingerprints; all differ across the matched seeds. \\
Requests per workload seed & Figure-specific. Each paired comparison keeps the same request count for all compared methods under the same workload seed; exact counts are stored in the corresponding archive. \\
Traffic profiles & Public-mixed explicit-QoS profiles, including concurrent-balanced, balanced-mixed, queue-focused, and safety-mixed variants; figure-specific variants are stored in the released archives. \\
SFC structures & Linear chains and SFC-DAGs with branches, merges, commutable segments, and dependency-preserving local order choices. \\
Instance concurrency & Configuration-specific; heavier or stateful VNF types use smaller concurrency limits than lightweight service types unless the snapshot specifies otherwise. \\
Queue estimator & Instance waiting proxy \(\widehat{W}_s(t)=Q_s(t)/\mu_s\), computed from observable queue length and service state. Main matched experiments use \(\epsilon_q=0\) to isolate scheduler effects; Appendix~\ref{app:queue_error_results} injects bounded nonzero errors and evaluates margin-protected validation. \\
\hline
\end{tabular}
\end{table*}

\begin{table*}[!tbp]
\caption{Orchestration and Evaluation Parameters}
\label{tab:orchestration_parameters}
\centering
\scriptsize
\setlength{\tabcolsep}{3pt}
\renewcommand{\arraystretch}{0.94}
\begin{tabular}{p{0.24\textwidth}p{0.68\textwidth}}
\hline
\textbf{Parameter} & \textbf{Value} \\
\hline
Access states & accept, trial, reject \\
Supported deployment profiles & conservative, balanced, throughput \\
Reported profile & The matched benchmark runs use the throughput profile unless otherwise stated. \\
Trial access & Borderline requests enter the same order/placement path with profile-specific screening or margins. \\
Top-$K$ candidate size & Runtime default $K=3$ unless explicitly overridden. Appendix sensitivity and selected control runs set $K=5$ where noted. \\
Candidate-order action set & Base order plus conservative linear-adjacent or directed branch-local serial rewrites; compact local-order action set with no sibling serialization. \\
Candidate scoring terms & Projected CPU pressure, queue waiting, path latency, service-pool reuse, critical/tail exposure, branch concentration, and future candidate scarcity. \\
Resource feasibility & Request-local CPU ledger with effective increments $\Delta r^{\rm cpu}_{v,s}$; global resource state advances only after a successful request result. \\
Path construction & Link-reachability and latency-aware path construction over the current snapshot. A separate universal residual-bandwidth hard guarantee is not part of the reported claim. \\
Queue treatment & Queue state enters order preview, instance ranking, and queue-aware end-to-end delay. No separate queue-capacity hard constraint is claimed. \\
Queue-estimation margin & 0 ms in the main matched runs. Appendix~\ref{app:queue_error} and the main-text queue-error robustness figure evaluate nonzero bounded-error margins. \\
Learning-module online role & Reorder retained top-$K$ instance candidates only; request-side order selection remains driven by deterministic placement-aware preview. \\
Successful state update & Service availability, request-local available CPU capacity, and queue-aware QoS are checked in the reported path; failed requests leave the shared runtime state unchanged. \\
Fig.~4 metrics & Per-request decision time, CPU imbalance, peak CPU, hot-node ratio, and queue delay. \\
Fig.~5 metrics & Variant gains relative to Neither and signed order--queue interaction for QoS ratio, queue delay, end-to-end delay, CPU imbalance, peak CPU, and runtime. \\
Fig.~6 metrics & Mechanism trace intensity and request-level QoS--resource outcomes, including learned re-ranking, QoS guards, QoS-boundary requests, CPU-pressure requests, and QoS failures. \\
\hline
\end{tabular}
\end{table*}

\begin{table*}[!tbp]
\caption{Learning-Assisted Re-Ranking Runtime Configuration}
\label{tab:learning_configuration}
\centering
\scriptsize
\setlength{\tabcolsep}{3pt}
\renewcommand{\arraystretch}{0.96}
\begin{tabular}{p{0.24\textwidth}p{0.68\textwidth}}
\hline
\textbf{Parameter} & \textbf{Value} \\
\hline
Online role & Revise the checking order inside a retained instance-candidate list; it refines candidate checking order before validation. \\
Runtime artifact & Frozen V5 ranking artifact loaded through the structure-aware ranking path in the reported runs. The main claim is limited to its online candidate-ordering role. \\
Action space & Masked categorical index over $\mathcal T^K_{i,v}$; $K$ is configuration-specific and the runtime default is 3. \\
Request/DAG inputs & Request class, QoS context, DAG size and branch/merge summaries, critical/tail labels, and current placement stage. \\
Candidate inputs & Deterministic score components, queue and resource state, path cost, remaining-capacity signal, reuse signal, tail risk, and future scarcity. \\
Output & Candidate index/order inside the retained list. \\
Request-side boundary & Legal local-order actions and placement-aware action preview are generated and selected outside the learned candidate path. \\
Fallback & When an accepted learned order is unavailable, the deterministic candidate order is used. \\
State-update boundary & The selected candidate continues through the common request-local CPU and queue-aware completion path; global state advances only after success. \\
Trace fields & Candidate list, deterministic order, learned suggestion, accepted change, guard/fallback events, and final request outcome. \\
Evaluation modes & Rule-only, shadow, learning-assisted re-ranking, and module-removal controls that preserve the deterministic score and common result path. \\
\hline
\end{tabular}
\end{table*}

\begin{table*}[!tbp]
\centering
\caption{Baseline fairness and reproduction scope. All methods are evaluated inside the same simulator with matched topology snapshots, workload traces, service-pool states, request counts, and seeds for each paired comparison.}
\label{tab:baseline_fairness}
\scriptsize
\setlength{\tabcolsep}{1.5pt}
\renewcommand{\arraystretch}{1.04}
\begin{tabular}{p{0.11\linewidth}p{0.20\linewidth}p{0.15\linewidth}p{0.15\linewidth}p{0.15\linewidth}p{0.16\linewidth}}
\toprule
Method & Input features & SFC-DAG and order handling & Queue / reuse use & Validator and fairness control & Training or reproduction scope \\
\midrule
Greedy-CPU & VNF type, node CPU state, feasible instance pool, and path feasibility. & Accepts SFC-DAG requests through the same request parser; uses the default legal order and no learned order action. & Queue and warm/cold status are visible to the common validator; the greedy score itself prioritizes remaining CPU. & Same shared success/evaluation path after candidate selection; the reported hard request-local checks are available CPU capacity and queue-aware QoS. & Self-reproduced heuristic; no training. Uses the same workload seeds as other methods. \\
Latency-Aware & Processing delay, link delay, path cost, candidate feasibility, and QoS budget. & Accepts SFC-DAG requests with the default legal order; no local order-action search. & Queue is checked by the common validator; the main score is latency and path oriented. Warm candidates are available in the common service pool. & Same shared success/evaluation path and request traces. & Self-reproduced latency-oriented baseline based on delay-aware VNF placement principles; no training. \\
DRL-DDQN & Normalized resource, queue, latency, QoS, and candidate-state summaries available before commitment. & Handles SFC-DAG requests after the same parser and default order abstraction; uses its own DDQN-style action interface rather than the RQ-SAFE local order-action set. & Uses the same observable queue/resource state in its state vector; warm reuse is available through the shared candidate pool. & Proposed actions are evaluated through the same request-result accounting and queue-aware QoS metrics. & Self-reproduced DDQN-style baseline; trained on the training split only, with held-out test workloads and archived random seeds. \\
GNN-DAG & Service-graph structure, candidate/node features, resource state, path features, and queue-related runtime features supplied by the same environment. & Supports SFC-DAG graph scoring; it uses graph-based scoring while RQ-SAFE supplies semantic local-order swap actions through its request abstraction. & Uses the same runtime state features where applicable; warm/cold instance availability is from the common service pool. & Uses the same request-result accounting and exported CPU/queue/QoS metrics after graph-based scoring. & Self-reproduced graph-aware SFC-DAG baseline with archived model settings, training split, and evaluation seeds. \\
RQ-SAFE & Request, SFC-DAG summaries, queue/resource state, top-\(K\) candidates, path cost, warm reuse, tail risk, and future scarcity features. & Supports SFC-DAG order abstraction and bounded legal local-order actions. & Queue state and warm reuse are explicit scoring and re-ranking features. & Same shared result path; learning can only reorder retained instance candidates before the request-local completion checks. & Frozen offline-trained ranking artifact; the online role and trace fields are summarized in Table~\ref{tab:learning_configuration}. \\
\bottomrule
\end{tabular}
\end{table*}

\section{Implementation Scope Clarifications}
\label{app:implementation_scope}
This section clarifies several implementation-sensitive components that are abstracted in the main text. The purpose is to state how these quantities enter candidate scoring, request-local completion checks, and the successful state-update path.

\subsection{Queue Estimator Scope and Margin Selection}
The FCFS proxy
\[
\widehat{W}_s(t)=Q_s(t)/\mu_s
\]
is a lightweight online queue-pressure estimator based on queue length and effective service rate. It keeps access control and validation inexpensive while exposing instance-level congestion to the scheduler. Its error can increase when service times are heterogeneous, arrivals are bursty, background tasks share the instance, the service rate changes because of co-location, or the runtime scheduler is not FCFS. The theoretical and experimental robustness analysis therefore treats queue delay as a bounded-error estimate and calibrates margins from validation residuals.

A practical deployment can choose the margin from recent validation residuals. Let
\[
e_{s,j}=W^{\mathrm{obs}}_{s,j}-\widehat{W}_{s,j}
\]
be the residual between observed waiting time and the online estimate for recent completed requests on instance \(s\). A profile-specific margin may be set as
\[
\epsilon_q(p)=\operatorname{Quantile}_{\rho_p}\left(|e_{s,j}|\right),
\]
where \(\rho_p\) is larger for conservative profiles and smaller for throughput-oriented profiles. The main reported runs set \(\epsilon_q=0\) to keep the scheduler comparison matched, while Appendix the main-text queue-error robustness figure injects \(1,3,5\) ms errors to show how nonzero margins trade access-control aggressiveness for QoS protection.

\subsection{Effective Resource Increment}
The effective resource increment \(\Delta r^r_{v,s}\) is the amount added to the temporary validation ledger if VNF \(v\) is assigned to instance \(s\). For CPU in the reported experiments, the raw demand \(r^{\mathrm{cpu}}_v\) is read from the benchmark VNF configuration. For a cold activation, the effective increment is the raw activation demand. For a warm or reusable instance, the increment is the additional task pressure on the already active instance, represented as a conservative reuse-adjusted demand,
\[
\Delta r^{\mathrm{cpu}}_{v,s}=\phi_{v,s}(t)r^{\mathrm{cpu}}_v,
\quad 0<\phi_{v,s}(t)\le 1,
\]
with implementation floors to avoid treating reuse as free. Thus, \(\Delta r^{\mathrm{cpu}}_{v,s}\) is not produced by the learned candidate ranker; it is a validation-side accounting quantity derived from the VNF demand, warm-instance status, reusable-instance evidence, and deployment profile. If production telemetry is available, \(\phi_{v,s}(t)\) can be calibrated from measured per-instance marginal CPU usage; in the submitted experiments it is determined by the benchmark configuration and conservative reuse rules.

\subsection{Implemented Semantic Compatibility and Structural Guards}
The released implementation uses explicit allow lists together with structural guards. The linear rule applies only to selected internal adjacent pairs. The DAG rule applies only to selected directed serial pairs on one branch with unique local predecessor/successor context. Merge and tail nodes are excluded. Independent siblings are not semantic swap actions. Table~\ref{tab:implemented_order_rules} records the actual rule families used by the main model.

A candidate rewrite is rejected when its pair is undeclared, its local context is ambiguous, or the pair crosses a protected merge/tail boundary. These conditions are deliberately sufficient rather than necessary. New service taxonomies require an explicit metadata/rule update before additional rewrites are enabled.

\subsection{Access-Control and Trial-Gate Parameters}
The risk-aware access-control rule uses a profile-dependent linear score. In the implementation, the equivalent form is
\[
S_i=w_{\mathrm{slack}}\,\mathrm{slack}_i
-w_{\mathrm{queue}}\,\mathrm{risk}^{q}_i
-w_{\mathrm{res}}\,\mathrm{cost}^{r}_i
-w_{\mathrm{len}}\,|\mathcal{V}_i|.
\]
The reported implementation uses the access-control weights in Table~\ref{tab:access_control_weights}. They were selected on validation workloads and then kept fixed for the reported test runs. The hard accept/reject thresholds are profile- and QoS-class-specific guards that act before final validation; borderline requests can still enter trial placement, but final commitment is controlled by the enabled deterministic checks, including the temporary CPU ledger and queue-aware QoS validation.

\begin{table}[!t]
\centering
\caption{Access-control score weights used by the reported implementation.}
\label{tab:access_control_weights}
\scriptsize
\setlength{\tabcolsep}{3pt}
\renewcommand{\arraystretch}{0.95}
\begin{tabular}{lrrrr}
\toprule
Profile & \(w_{\mathrm{slack}}\) & \(w_{\mathrm{queue}}\) & \(w_{\mathrm{res}}\) & \(w_{\mathrm{len}}\) \\
\midrule
Light-load / conservative & 0.50 & 0.90 & 7.0 & 0.7 \\
Default / balanced & 0.45 & 1.00 & 8.0 & 0.8 \\
Heavy-load / throughput-stress & 0.40 & 1.20 & 9.5 & 1.0 \\
\bottomrule
\end{tabular}
\end{table}

\subsection{Learning-Assisted Runtime Boundary}
The learned component receives candidate-level pre-commit features and returns a revised order inside the retained list. Reward definitions, optimizer choices, and training schedules are not part of the online orchestration model claimed by the paper. The released evaluation uses a frozen artifact. Candidate feasibility and request-result accounting remain in the common runtime path described in Appendix~\ref{app:learning_reranking}.

\subsection{Path Backtracking and Termination}
Path repair/backtracking is bounded by the retained local candidate list and by the retained local order-action set. The implementation searches retained alternatives rather than all complete infrastructure mappings. If the current candidate fails a service-availability, CPU-capacity, configured-path, or QoS check, the validator tries the next retained candidate for the affected stage and may trigger a new configured path search. Let \(K_\pi\) be the retained order count and let \(B\le K\) be the maximum number of candidate retries per VNF and per order. The corresponding upper bound is given in Appendix~\ref{app:complexity}. Appendix the main-text top-$K$ sensitivity figure and Table~\ref{tab:topology_method_summary} report the observed per-request decision-time tradeoff.

\begin{table*}[!tbp]
\centering
\caption{Implementation-sensitive component boundaries.}
\label{tab:scope_clarification_summary}
\scriptsize
\setlength{\tabcolsep}{1.5pt}
\renewcommand{\arraystretch}{1.02}
\begin{tabular}{p{0.18\linewidth}p{0.33\linewidth}p{0.22\linewidth}p{0.19\linewidth}}
\toprule
Component & Reported interpretation & Deployment consideration & Supporting evidence \\
\midrule
Queue estimate & Lightweight instance waiting estimate used in preview, candidate ranking, and queue-aware delay. & Calibrate a nonzero margin when production residuals are material. & Proposition A.3 and the main-text queue-error robustness figure. \\
Effective CPU increment & Request-local ledger increment that distinguishes cold activation from conservative warm reuse. & Recalibrate marginal CPU when telemetry differs from benchmark profiles. & Proposition A.1 and runtime control table. \\
Local-order semantics & Explicit linear-adjacent and branch-local serial rule families; unknown pairs are disabled. & New VNF types require declared semantic and structural rules. & Table~\ref{tab:implemented_order_rules}. \\
Path handling & Reachability and link latency participate in candidate/path construction. & Add residual-bandwidth reservation if the deployment requires a bandwidth guarantee. & Tables~\ref{tab:simulation_settings} and~\ref{tab:orchestration_parameters}. \\
Learning-assisted re-ranking & Frozen artifact revises only the retained candidate order. & Artifact and gate can be recalibrated without changing the semantic action or completion path. & Table~\ref{tab:learning_configuration} and Proposition C.1. \\
Fallback & Finite retry over retained instance candidates and retained legal actions. & Wider retained sets increase decision time. & Complexity analysis, the main-text top-$K$ sensitivity figure, and Table~\ref{tab:app_order_budget_kpi}. \\
\bottomrule
\end{tabular}
\end{table*}

\section{Additional Sensitivity and Robustness Evidence}
\label{app:additional_figures}

This section provides supplementary evidence on retained-candidate sensitivity, retained order-action budget, queue-estimation error, runtime controls, decision-time decomposition, and profile stress. The top-$K$ sensitivity and queue-error robustness figures are placed in the main text, while the tables below give detailed numerical settings and controls.

\subsection{Top-$K$ Candidate-Width Sensitivity}
The top-$K$ sensitivity figure in the main text evaluates the retained candidate width $K$. The result supports using a moderate candidate width rather than an overly narrow search-limited list or an over-wide list with extra online cost.

\subsection{Queue-Estimation-Error Robustness}
\label{app:queue_error_results}
The queue-estimation-error robustness figure in the main text reports the queue-estimation-error robustness study. All values are relative to the exact queue-estimate baseline. Under-estimating the queue without a margin can reduce actual QoS compliance, whereas adding the validation margin $|\mathcal{V}_i|\epsilon_q$ moves the decision into a more conservative region and recovers QoS at the cost of lower acceptance/success. This supports the bounded-error argument in Appendix~\ref{app:feasibility_qos}: queue margins protect QoS by trading off conservativeness.

\subsection{Runtime Control Boundary}
Table~\ref{tab:appendix_guard_runtime_check} summarizes the implemented boundary used by the main-text claims.

\begin{table*}[!tbp]
\centering
\caption{Runtime control boundary of RQ-SAFE.}
\label{tab:appendix_guard_runtime_check}
\small
\begin{tabular}{p{0.22\linewidth}p{0.36\linewidth}p{0.34\linewidth}}
\hline
\textbf{Component} & \textbf{Implemented path} & \textbf{Claim used in the paper} \\
\hline
Service and CPU availability & Candidate construction uses available service instances and request-local projected CPU increments. & A candidate without service availability or available CPU capacity is skipped. \\
Queue-inclusive QoS & Complete-plan latency is recomputed with instance waiting estimates and the request budget. & Successful results satisfy the configured queue-aware QoS test. \\
State update & Batch replay updates shared CPU and queue state only after a successful request result. & Failed requests do not advance the shared runtime state. \\
Path handling & Current path reachability/link latency enter candidate construction. & The paper claims path-aware candidate construction, not a universal residual-bandwidth guarantee. \\
Learning action & The learned output is an order/index inside a retained candidate list. & Learning refines candidate order within the retained list; semantic order actions are selected by the explicit request-side mechanism. \\
Queue-error margin & Nonzero error and margin fields are exercised by the robustness experiment. & Bounded queue error can be traded against access-control aggressiveness through a validation margin. \\
\hline
\end{tabular}
\end{table*}

\subsection{Profile Stress Under Queue Under-Estimation}
Table~\ref{tab:app_profile_stress_replay} reports the compact profile-control experiment moved from the main text. The table separates validation-side decisions under a 15\% underestimated queue contribution from recovered QoS under the injected true contribution. Conservative profiles preserve a smaller accepted set; balanced and throughput profiles admit more requests but lose recovered QoS. The learning-assisted and rule-only rows show that the profile trend is not created solely by candidate re-ranking.

\begin{table}[!t]
    \centering
    \caption{Profile-stress replay with learning-assisted/rule-only controls under 15\% queue under-estimation.}
    \label{tab:app_profile_stress_replay}
    \tiny
    \setlength{\tabcolsep}{1.6pt}
    \begin{tabular}{llrrrr}
        \hline
        Mode & Profile & Val. QoS (\%) & Rec. QoS (\%) & Rec. budget (\%) & Decision (ms) \\
        \hline
        Learned & Cons. & 74.46 & 74.46 & 78.03 & 611.58 \\
        Learned & Bal. & 80.43 & 72.84 & 78.21 & 616.17 \\
        Learned & Thr. & 82.13 & 72.84 & 78.21 & 618.11 \\
        Rule-only & Cons. & 74.17 & 74.17 & 77.94 & 505.02 \\
        Rule-only & Bal. & 80.89 & 73.34 & 78.11 & 509.40 \\
        Rule-only & Thr. & 82.34 & 72.68 & 78.28 & 512.02 \\
        \hline
    \end{tabular}
\end{table}

\subsection{Runtime Traceability as Secondary Evidence}

The runtime traceability figure in the main text summarizes this evidence. Final validation is recorded for every request in the reported set. Accepted learning-assisted re-ranking appears at $29.59$ events per 100 requests, while guard records are more frequent because a single request can produce several candidate-level checks. Access-risk and DAG/order values count intermediate trace records and therefore use a different denominator and interpretation.

The request map contains 268 outcomes, including 79 near the QoS boundary, 12 in the CPU-pressure region, and 26 QoS failures. Circled points identify requests for which an accepted candidate re-ranking is present in the recorded path. The figure establishes observability of actions, candidates, guards, fallback, and outcomes. Performance conclusions continue to rely on the matched comparisons and factorial experiment rather than on trace counts.

Table~\ref{tab:app_request_trace_example} gives one decoded request path from the recorded log. It illustrates how local-order exposure, queue-aware preview, candidate construction, and final queue-aware validation are connected in a single inspectable record.

\begin{table*}[!t]
\centering
\caption{Decoded request-level trace used as an example of the recorded decision path.}
\label{tab:app_request_trace_example}
\scriptsize
\setlength{\tabcolsep}{3pt}
\renewcommand{\arraystretch}{1.03}
\begin{tabular}{p{0.17\textwidth}p{0.40\textwidth}p{0.35\textwidth}}
\hline
Trace step & Recorded values & Interpretation \\
\hline
Input & \texttt{req\_045}; six VNFs; general-DAG low-latency request; legal serial pair $(v_2,v_3)$; QoS budget $52.75$ ms. & The request has bounded local-order freedom and an explicit end-to-end budget. \\
Access control & Estimated queue latency $15.83$ ms; QoS slack $36.92$ ms; queue-risk score $3.456$; access decision \emph{accept}. & Queue state is visible before placement, but commitment is not made at access control. \\
Order preview & Base order filter--detect--compress--cache--aggregate--encrypt has adjusted preview latency $41.30$ ms. The rewritten order filter--compress--detect--cache--aggregate--encrypt has adjusted preview latency $36.30$ ms. Queue preview changes from $16.59$ to $16.06$ ms. & The legal rewrite is selected because the provisional placement improves adjusted latency by $4.995$ ms. \\
Candidate scoring & $56$ candidates are scored and $16$ are filtered as unsupported or unavailable. The selected placement is node-a, node-c, node-b, node-d, node-b, node-c for the six VNFs. & Candidate construction is conditioned on the retained order and service-pool state. \\
Validation and result & Final queue-aware adjusted delay is $40.77$ ms, below the $52.75$ ms budget; queue delay is $16.06$ ms; QoS passes; recorded decision time is $986.83$ ms. & The request becomes a committed success only after complete-plan validation. \\
\hline
\end{tabular}
\end{table*}

\subsection{Additional Factorial Details}
Table~\ref{tab:app_workload_interaction} decomposes the order--queue interaction by workload and includes the metrics omitted from the main-text summary. The QoS interaction remains positive in all four workload families. The queue-delay, end-to-end-delay, CPU, and peak-CPU interactions are mixed, which is why the main text interprets non-additive coupling most strongly for QoS-compliant service outcome.

\begin{table*}[!t]
\centering
\caption{Workload-level order--queue interaction. Values are signed gains; positive is preferable for each metric.}
\label{tab:app_workload_interaction}
\scriptsize
\setlength{\tabcolsep}{3.2pt}
\begin{tabular}{lrrrrrr}
\hline
Workload & QoS & Queue & E2E & CPU imb. & Peak CPU & Decision \\
 & (pp) & (ms) & (ms) & (pp) & (pp) & (ms) \\
\hline
W1 Concurrent-balanced & $+4.69$ & $-0.82$ & $-0.27$ & $+2.46$ & $+1.92$ & $+15.90$ \\
W2 Balanced-mixed & $+7.29$ & $+0.03$ & $+1.09$ & $+2.40$ & $-5.60$ & $+50.27$ \\
W3 Queue-focused & $+0.95$ & $-0.87$ & $+0.07$ & $-3.63$ & $-0.28$ & $+12.30$ \\
W4 Safety-mixed & $+2.38$ & $-0.55$ & $-0.24$ & $+0.11$ & $-0.56$ & $+64.77$ \\
Overall & $+3.83$ & $-0.55$ & $+0.16$ & $+0.34$ & $-1.13$ & $+35.81$ \\
\hline
\end{tabular}
\end{table*}

\subsection{Stage-3 Checks: Retained Order-Action Budget and Decision Time}
\label{app:stage3_checks}

Table~\ref{tab:app_order_budget_kpi} reports the retained order-action budget check. The action-rich suite contains 4 workload families, 3 seeds, and 4 variants for each budget value, giving 192 raw run rows. Each order-enabled variant contains 804 requests across the matched workload--seed groups. In this suite, every eligible request exposes the default order and one non-default local rewrite. Hence $K_\pi=1$ disables the non-default rewrite under the same order-enabled code path, and $K_\pi\ge2$ saturates the retained order-action set. The table should therefore be read as a saturation check for the retained action set, not as a claim that larger order budgets are always redundant in workloads with multiple legal local rewrites.

\begin{table*}[!t]
\centering
\caption{Retained order-action budget check in the action-rich suite. Adoption is computed over retained eligible requests. QoS is the QoS-compliant service ratio, and time is mean per-request decision time.}
\label{tab:app_order_budget_kpi}
\scriptsize
\setlength{\tabcolsep}{3.2pt}
\renewcommand{\arraystretch}{0.98}
\begin{tabular}{crrrrrrr}
\hline
$K_\pi$ & Variant & Requests & Retained eligible & Applied non-default & Adoption & QoS & Time \\
 & & & & & (\%) & (\%) & (ms) \\
\hline
1 & Order only & 804 & 0 & 0 & -- & 53.76 & 380.56 \\
1 & Coupled & 804 & 0 & 0 & -- & 54.53 & 368.36 \\
2 & Order only & 804 & 804 & 221 & 27.49 & 53.69 & 957.01 \\
2 & Coupled & 804 & 804 & 336 & 41.79 & 58.29 & 928.16 \\
3 & Order only & 804 & 804 & 221 & 27.49 & 53.69 & 953.19 \\
3 & Coupled & 804 & 804 & 336 & 41.79 & 58.29 & 943.12 \\
5 & Order only & 804 & 804 & 221 & 27.49 & 53.69 & 968.50 \\
5 & Coupled & 804 & 804 & 336 & 41.79 & 58.29 & 931.26 \\
\hline
\end{tabular}
\end{table*}

The results show two effects. First, exposing the first non-default local rewrite activates the order-side mechanism: the coupled variant selects a non-default local order for 336 of 804 requests, compared with 221 of 804 in the order-only variant. Second, the $K_\pi=2,3,5$ rows are identical in service and action outcomes because the suite exposes only one non-default local action per eligible request. The coupled QoS ratio increases from 54.53\% at $K_\pi=1$ to 58.29\% once the retained rewrite is available, while the order-only QoS ratio is almost unchanged. This supports the interpretation that order freedom becomes useful when it is evaluated through current queue/resource state.

Table~\ref{tab:app_decision_time_breakdown} decomposes the recorded decision time. The labels below use the paper-level terminology. The implementation timing field named access control corresponds to access-control screening in the paper. The decomposition separates request normalization, semantic order abstraction, access-control screening, placement-and-validation search, QoS evaluation, queue-state evaluation, and residual runtime/logging cost.

\begin{table*}[!t]
\centering
\caption{Per-request decision-time decomposition from 804 request-level timing records.}
\label{tab:app_decision_time_breakdown}
\scriptsize
\setlength{\tabcolsep}{4pt}
\renewcommand{\arraystretch}{0.98}
\begin{tabular}{lrrrr}
\hline
Paper-level stage & Mean (ms) & Median (ms) & 95th perc. (ms) & Share of mean total (\%) \\
\hline
Request translation/normalization & 47.47 & 47.89 & 68.45 & 12.58 \\
Semantic order abstraction & 0.07 & 0.05 & 0.12 & 0.02 \\
Access-control screening & 1.22 & 1.15 & 1.86 & 0.32 \\
Placement-and-validation search & 256.01 & 250.80 & 360.86 & 67.84 \\
QoS evaluation & 27.29 & 26.03 & 34.64 & 7.23 \\
Queue-state evaluation & 0.07 & 0.06 & 0.12 & 0.02 \\
Residual runtime/logging/serialization & 45.24 & 39.95 & 84.05 & 11.99 \\
\hline
\end{tabular}
\end{table*}

The decomposition shows that the observed decision-time cost is dominated by placement-and-validation search. Semantic order abstraction and access-control screening together account for less than 0.4\% of the mean decision time. This supports the main-text interpretation that the additional runtime cost mainly comes from checked candidate construction and validation over retained alternatives, rather than from merely extracting local-order actions.

\subsection{Claim-to-Evidence Map}
Table~\ref{tab:claim_evidence_map} links the main claims to the supporting appendix evidence.

\begin{table*}[!tbp]
\centering
\caption{Main-text claim to appendix-evidence map.}
\label{tab:claim_evidence_map}
\scriptsize
\setlength{\tabcolsep}{3pt}
\renewcommand{\arraystretch}{1.02}
\begin{tabular}{p{0.24\linewidth}p{0.24\linewidth}p{0.29\linewidth}p{0.17\linewidth}}
\toprule
Main-text claim & Evidence & Supported interpretation & Scope \\
\midrule
Request-side freedom is a bounded operational variable. & Appendix~\ref{app:order_preview}, Table~\ref{tab:implemented_order_rules}, and Table~\ref{tab:app_order_budget_kpi}. & Only declared linear-adjacent or branch-local serial rewrites are exposed; siblings remain parallel, and the budget and factorial suites record how often the enabled rewrite is adopted. & Rule set is benchmark-taxonomy specific; the budget check saturates at one non-default rewrite in the reported action-rich suite. \\
Request and resource sides are coupled through placement consequences. & Placement-aware preview, recursive candidate model, main-text Fig.~5, and Tables~\ref{tab:app_request_trace_example} and~\ref{tab:app_workload_interaction}. & A legal action changes provisional placement; queue/resource feedback changes whether that action is retained and how later VNFs are placed. & The statistically clear non-additive effect is concentrated in QoS outcome. \\
Learning has bounded online influence. & Appendix~\ref{app:learning_reranking}, Table~\ref{tab:learning_configuration}, and Proposition C.1. & The artifact only reorders retained instance candidates and shares the common request-result path. & No new training-algorithm claim is made. \\
Queue-error margins trade aggressiveness for robustness. & Proposition A.3 and the main-text queue-error robustness figure. & A bounded per-stage error can be absorbed by a path-length/request-size margin. & Guarantee follows the bounded-error assumption. \\
Moderate retained width balances search and overhead. & the main-text top-$K$ sensitivity figure, Table~\ref{tab:app_order_budget_kpi}, Table~\ref{tab:app_decision_time_breakdown}, and complexity analysis. & Increasing retained candidate/order coverage can increase online cost; the measured cost is dominated by placement-and-validation search. & Optimal $K$ and $K_\pi$ are deployment specific. \\
Cross-method results use matched workload state. & Tables~\ref{tab:simulation_settings},~\ref{tab:baseline_fairness}, and~\ref{tab:topology_method_summary}. & Paired methods share topology, requests, seed, service-pool snapshot, and metric accounting. & Baseline implementations follow the stated information boundaries. \\
The pipeline remains competitive across topology scale. & Tables~\ref{tab:topology_method_summary} and~\ref{tab:appendix_raw_topology_runs}. & Strict replay covers three public topologies, four workload groups, five methods, and matched seeds. & Evidence is simulation/replay based. \\
\bottomrule
\end{tabular}
\end{table*}

\clearpage
\onecolumn
\section{Strict Topology-Scale Paired Comparison and Raw Run-Level Results}
\label{app:raw_topology_results}
Tables~\ref{tab:topology_method_summary} and~\ref{tab:appendix_raw_topology_runs} report the topology-scale replay used for the appendix comparison. The replay covers $3$ topologies $\times$ $4$ workload groups $\times$ $5$ methods $\times$ $3$ seeds, for a total of $180$ runs. Each topology--workload--seed group uses the same request trace for all methods. The completeness runtime check passes for all runs: each method is evaluated on the same $2,412$ arrivals, and every arrival contributes exactly one result row. This strict accounting removes the request-count ambiguity in the earlier export.

Under this strict replay, RQ-SAFE and GNN-DAG form the leading service-outcome group. Their success/QoS rates are close under strict accounting, while both methods clearly outperform Latency-Aware, DRL-DDQN, and Greedy-CPU. Therefore, this topology-scale replay supports paired-accounting robustness and competitiveness with the graph-aware baseline, together with distinct placement behavior for the two methods. A placement-signature runtime check also confirms that the two methods are not collapsed into the same decision path: final assigned-node signatures differ in 2164 of 2412 paired requests.

\begin{table*}[!tbp]
\centering
\caption{Strict paired topology-scale summary.}
\label{tab:topology_method_summary}
\scriptsize
\setlength{\tabcolsep}{3.8pt}
\renewcommand{\arraystretch}{0.96}
\begin{tabular}{lrrrrrrrr}
\toprule
Method & Runs & Arrivals & Rows & Success & QoS & Queue & Budget & Decision \\
 & & & & (\%) & (\%) & (ms) & (\%) & (ms) \\
\midrule
RQ-SAFE & 36 & 2412 & 2412 & 74.6 & 74.6 & 22.0 & 77.8 & 654.6 \\
GNN-DAG & 36 & 2412 & 2412 & 74.9 & 74.9 & 21.8 & 77.3 & 616.0 \\
Latency-Aware & 36 & 2412 & 2412 & 64.1 & 64.1 & 38.5 & 93.7 & 154.3 \\
DRL-DDQN & 36 & 2412 & 2412 & 56.6 & 56.6 & 23.1 & 97.5 & 293.4 \\
Greedy-CPU & 36 & 2412 & 2412 & 51.4 & 51.4 & 31.9 & 101.9 & 149.8 \\
\bottomrule
\end{tabular}
\vspace{1mm}
\parbox{0.96\textwidth}{\footnotesize Notes: Arrivals is the number of replayed input requests. Rows is the number of request-level result rows after strict completeness filtering. All methods are evaluated on the same 2,412 arrivals. Success and QoS use the same denominator in this topology-scale replay, so they are equal in this table.}
\end{table*}

\begin{ThreePartTable}
\begin{TableNotes}[flushleft]
\footnotesize
\item Notes: Each row corresponds to one topology--workload--method--seed run. Arrivals and Rows verify one-to-one request accounting for the run. Success and QoS are percentages. Queue, budget, and decision time are run-level means over requests. Budget is the mean queue-aware delay-budget usage. The corresponding CSV files and completeness runtime check are archived with the experimental artifacts.
\end{TableNotes}
\scriptsize
\setlength{\tabcolsep}{2.0pt}
\renewcommand{\arraystretch}{0.90}
\setlength{\LTleft}{0pt}
\setlength{\LTright}{0pt}
\setlength{\LTcapwidth}{\linewidth}
\begin{longtable}{@{}>{\raggedright\arraybackslash}p{0.085\linewidth}>{\centering\arraybackslash}p{0.070\linewidth}>{\centering\arraybackslash}p{0.055\linewidth}>{\centering\arraybackslash}p{0.045\linewidth}>{\raggedright\arraybackslash}p{0.115\linewidth}>{\centering\arraybackslash}p{0.052\linewidth}>{\centering\arraybackslash}p{0.052\linewidth}>{\centering\arraybackslash}p{0.062\linewidth}>{\centering\arraybackslash}p{0.052\linewidth}>{\centering\arraybackslash}p{0.058\linewidth}>{\centering\arraybackslash}p{0.062\linewidth}>{\centering\arraybackslash}p{0.075\linewidth}@{}}
\caption{Complete strict run-level results for the topology-scale paired comparison.}\label{tab:appendix_raw_topology_runs}\\
\toprule
Topology & Size & Workload & Seed & Method & Arr. & Rows & Success & QoS & Queue & Budget & Decision \\
 & & & & & & & (\%) & (\%) & (ms) & (\%) & (ms) \\
\midrule
\endfirsthead
\caption[]{Complete strict run-level results for the topology-scale paired comparison (continued).}\\
\toprule
Topology & Size & Workload & Seed & Method & Arr. & Rows & Success & QoS & Queue & Budget & Decision \\
 & & & & & & & (\%) & (\%) & (ms) & (\%) & (ms) \\
\midrule
\endhead
\midrule
\multicolumn{12}{r}{Continued on next page}\\
\endfoot
\bottomrule
\insertTableNotes
\endlastfoot
Aarnet & 19N/48E & W1 & 831 & RQ-SAFE & 64 & 64 & 62.5 & 62.5 & 23.6 & 87.9 & 615.1 \\
Aarnet & 19N/48E & W1 & 831 & GNN-DAG & 64 & 64 & 62.5 & 62.5 & 23.5 & 88.5 & 488.7 \\
Aarnet & 19N/48E & W1 & 831 & Latency-Aware & 64 & 64 & 53.1 & 53.1 & 37.8 & 104.3 & 132.6 \\
Aarnet & 19N/48E & W1 & 831 & DRL-DDQN & 64 & 64 & 32.8 & 32.8 & 24.0 & 115.5 & 233.4 \\
Aarnet & 19N/48E & W1 & 831 & Greedy-CPU & 64 & 64 & 51.6 & 51.6 & 31.8 & 101.0 & 137.4 \\
Aarnet & 19N/48E & W1 & 832 & RQ-SAFE & 64 & 64 & 62.5 & 62.5 & 23.7 & 89.4 & 596.0 \\
Aarnet & 19N/48E & W1 & 832 & GNN-DAG & 64 & 64 & 62.5 & 62.5 & 23.6 & 88.6 & 507.2 \\
Aarnet & 19N/48E & W1 & 832 & Latency-Aware & 64 & 64 & 53.1 & 53.1 & 37.8 & 104.3 & 139.3 \\
Aarnet & 19N/48E & W1 & 832 & DRL-DDQN & 64 & 64 & 32.8 & 32.8 & 24.0 & 115.5 & 266.0 \\
Aarnet & 19N/48E & W1 & 832 & Greedy-CPU & 64 & 64 & 51.6 & 51.6 & 31.8 & 101.0 & 145.1 \\
Aarnet & 19N/48E & W1 & 833 & RQ-SAFE & 64 & 64 & 62.5 & 62.5 & 23.7 & 89.4 & 586.9 \\
Aarnet & 19N/48E & W1 & 833 & GNN-DAG & 64 & 64 & 62.5 & 62.5 & 23.6 & 88.4 & 517.2 \\
Aarnet & 19N/48E & W1 & 833 & Latency-Aware & 64 & 64 & 53.1 & 53.1 & 37.8 & 104.2 & 136.1 \\
Aarnet & 19N/48E & W1 & 833 & DRL-DDQN & 64 & 64 & 32.8 & 32.8 & 24.0 & 115.5 & 243.6 \\
Aarnet & 19N/48E & W1 & 833 & Greedy-CPU & 64 & 64 & 51.6 & 51.6 & 31.8 & 101.0 & 127.6 \\
Aarnet & 19N/48E & W2 & 831 & RQ-SAFE & 64 & 64 & 65.6 & 65.6 & 22.9 & 87.7 & 588.8 \\
Aarnet & 19N/48E & W2 & 831 & GNN-DAG & 64 & 64 & 65.6 & 65.6 & 22.9 & 84.6 & 514.5 \\
Aarnet & 19N/48E & W2 & 831 & Latency-Aware & 64 & 64 & 57.8 & 57.8 & 37.9 & 103.3 & 161.4 \\
Aarnet & 19N/48E & W2 & 831 & DRL-DDQN & 64 & 64 & 40.6 & 40.6 & 23.6 & 111.3 & 271.4 \\
Aarnet & 19N/48E & W2 & 831 & Greedy-CPU & 64 & 64 & 50.0 & 50.0 & 32.4 & 100.6 & 142.4 \\
Aarnet & 19N/48E & W2 & 832 & RQ-SAFE & 64 & 64 & 65.6 & 65.6 & 22.9 & 87.9 & 599.4 \\
Aarnet & 19N/48E & W2 & 832 & GNN-DAG & 64 & 64 & 65.6 & 65.6 & 23.0 & 86.1 & 557.6 \\
Aarnet & 19N/48E & W2 & 832 & Latency-Aware & 64 & 64 & 57.8 & 57.8 & 37.9 & 103.2 & 141.6 \\
Aarnet & 19N/48E & W2 & 832 & DRL-DDQN & 64 & 64 & 40.6 & 40.6 & 23.6 & 111.2 & 259.9 \\
Aarnet & 19N/48E & W2 & 832 & Greedy-CPU & 64 & 64 & 50.0 & 50.0 & 32.4 & 100.6 & 134.6 \\
Aarnet & 19N/48E & W2 & 833 & RQ-SAFE & 64 & 64 & 65.6 & 65.6 & 22.9 & 87.9 & 601.9 \\
Aarnet & 19N/48E & W2 & 833 & GNN-DAG & 64 & 64 & 65.6 & 65.6 & 22.9 & 84.4 & 541.0 \\
Aarnet & 19N/48E & W2 & 833 & Latency-Aware & 64 & 64 & 57.8 & 57.8 & 37.9 & 103.3 & 133.1 \\
Aarnet & 19N/48E & W2 & 833 & DRL-DDQN & 64 & 64 & 40.6 & 40.6 & 23.6 & 111.1 & 252.3 \\
Aarnet & 19N/48E & W2 & 833 & Greedy-CPU & 64 & 64 & 50.0 & 50.0 & 32.4 & 100.6 & 140.9 \\
Aarnet & 19N/48E & W3 & 831 & RQ-SAFE & 70 & 70 & 71.4 & 71.4 & 20.4 & 82.0 & 586.8 \\
Aarnet & 19N/48E & W3 & 831 & GNN-DAG & 70 & 70 & 77.1 & 77.1 & 20.5 & 80.1 & 542.7 \\
Aarnet & 19N/48E & W3 & 831 & Latency-Aware & 70 & 70 & 64.3 & 64.3 & 37.3 & 94.9 & 149.5 \\
Aarnet & 19N/48E & W3 & 831 & DRL-DDQN & 70 & 70 & 45.7 & 45.7 & 21.1 & 103.1 & 275.1 \\
Aarnet & 19N/48E & W3 & 831 & Greedy-CPU & 70 & 70 & 60.0 & 60.0 & 34.1 & 98.4 & 152.3 \\
Aarnet & 19N/48E & W3 & 832 & RQ-SAFE & 70 & 70 & 70.0 & 70.0 & 20.3 & 82.3 & 601.7 \\
Aarnet & 19N/48E & W3 & 832 & GNN-DAG & 70 & 70 & 75.7 & 75.7 & 20.5 & 79.4 & 527.7 \\
Aarnet & 19N/48E & W3 & 832 & Latency-Aware & 70 & 70 & 64.3 & 64.3 & 37.2 & 94.8 & 177.9 \\
Aarnet & 19N/48E & W3 & 832 & DRL-DDQN & 70 & 70 & 45.7 & 45.7 & 21.2 & 103.1 & 271.1 \\
Aarnet & 19N/48E & W3 & 832 & Greedy-CPU & 70 & 70 & 61.4 & 61.4 & 34.2 & 98.6 & 153.3 \\
Aarnet & 19N/48E & W3 & 833 & RQ-SAFE & 70 & 70 & 70.0 & 70.0 & 20.3 & 81.9 & 575.5 \\
Aarnet & 19N/48E & W3 & 833 & GNN-DAG & 70 & 70 & 75.7 & 75.7 & 20.5 & 80.4 & 504.4 \\
Aarnet & 19N/48E & W3 & 833 & Latency-Aware & 70 & 70 & 64.3 & 64.3 & 37.3 & 94.9 & 146.6 \\
Aarnet & 19N/48E & W3 & 833 & DRL-DDQN & 70 & 70 & 44.3 & 44.3 & 21.1 & 103.1 & 258.3 \\
Aarnet & 19N/48E & W3 & 833 & Greedy-CPU & 70 & 70 & 60.0 & 60.0 & 34.2 & 98.6 & 137.0 \\
Aarnet & 19N/48E & W4 & 831 & RQ-SAFE & 70 & 70 & 75.7 & 75.7 & 21.1 & 81.6 & 529.2 \\
Aarnet & 19N/48E & W4 & 831 & GNN-DAG & 70 & 70 & 74.3 & 74.3 & 20.9 & 80.6 & 485.5 \\
Aarnet & 19N/48E & W4 & 831 & Latency-Aware & 70 & 70 & 64.3 & 64.3 & 36.8 & 96.6 & 136.8 \\
Aarnet & 19N/48E & W4 & 831 & DRL-DDQN & 70 & 70 & 38.6 & 38.6 & 21.3 & 105.8 & 261.1 \\
Aarnet & 19N/48E & W4 & 831 & Greedy-CPU & 70 & 70 & 58.6 & 58.6 & 32.8 & 98.9 & 155.9 \\
Aarnet & 19N/48E & W4 & 832 & RQ-SAFE & 70 & 70 & 75.7 & 75.7 & 21.1 & 81.4 & 635.7 \\
Aarnet & 19N/48E & W4 & 832 & GNN-DAG & 70 & 70 & 74.3 & 74.3 & 20.9 & 81.3 & 497.6 \\
Aarnet & 19N/48E & W4 & 832 & Latency-Aware & 70 & 70 & 64.3 & 64.3 & 36.8 & 96.7 & 143.1 \\
Aarnet & 19N/48E & W4 & 832 & DRL-DDQN & 70 & 70 & 40.0 & 40.0 & 21.4 & 106.0 & 275.3 \\
Aarnet & 19N/48E & W4 & 832 & Greedy-CPU & 70 & 70 & 57.1 & 57.1 & 32.7 & 98.8 & 140.6 \\
Aarnet & 19N/48E & W4 & 833 & RQ-SAFE & 70 & 70 & 74.3 & 74.3 & 21.1 & 81.6 & 588.1 \\
Aarnet & 19N/48E & W4 & 833 & GNN-DAG & 70 & 70 & 77.1 & 77.1 & 20.9 & 80.3 & 517.4 \\
Aarnet & 19N/48E & W4 & 833 & Latency-Aware & 70 & 70 & 64.3 & 64.3 & 36.8 & 96.6 & 130.9 \\
Aarnet & 19N/48E & W4 & 833 & DRL-DDQN & 70 & 70 & 41.4 & 41.4 & 21.4 & 106.0 & 252.4 \\
Aarnet & 19N/48E & W4 & 833 & Greedy-CPU & 70 & 70 & 58.6 & 58.6 & 32.7 & 98.9 & 130.8 \\
\addlinespace[1.2pt]
Abvt & 23N/62E & W1 & 831 & RQ-SAFE & 64 & 64 & 70.3 & 70.3 & 23.6 & 78.4 & 632.9 \\
Abvt & 23N/62E & W1 & 831 & GNN-DAG & 64 & 64 & 68.8 & 68.8 & 23.3 & 79.1 & 580.6 \\
Abvt & 23N/62E & W1 & 831 & Latency-Aware & 64 & 64 & 59.4 & 59.4 & 39.9 & 95.9 & 182.9 \\
Abvt & 23N/62E & W1 & 831 & DRL-DDQN & 64 & 64 & 56.2 & 56.2 & 24.5 & 99.6 & 351.6 \\
Abvt & 23N/62E & W1 & 831 & Greedy-CPU & 64 & 64 & 48.4 & 48.4 & 31.3 & 103.9 & 165.7 \\
Abvt & 23N/62E & W1 & 832 & RQ-SAFE & 64 & 64 & 70.3 & 70.3 & 23.6 & 79.1 & 791.0 \\
Abvt & 23N/62E & W1 & 832 & GNN-DAG & 64 & 64 & 70.3 & 70.3 & 23.3 & 79.0 & 600.6 \\
Abvt & 23N/62E & W1 & 832 & Latency-Aware & 64 & 64 & 59.4 & 59.4 & 39.9 & 95.9 & 148.7 \\
Abvt & 23N/62E & W1 & 832 & DRL-DDQN & 64 & 64 & 56.2 & 56.2 & 24.5 & 99.6 & 299.3 \\
Abvt & 23N/62E & W1 & 832 & Greedy-CPU & 64 & 64 & 48.4 & 48.4 & 31.3 & 103.9 & 143.7 \\
Abvt & 23N/62E & W1 & 833 & RQ-SAFE & 64 & 64 & 70.3 & 70.3 & 23.4 & 80.0 & 719.3 \\
Abvt & 23N/62E & W1 & 833 & GNN-DAG & 64 & 64 & 68.8 & 68.8 & 23.3 & 79.1 & 589.6 \\
Abvt & 23N/62E & W1 & 833 & Latency-Aware & 64 & 64 & 59.4 & 59.4 & 39.9 & 95.9 & 170.8 \\
Abvt & 23N/62E & W1 & 833 & DRL-DDQN & 64 & 64 & 56.2 & 56.2 & 24.5 & 99.6 & 326.7 \\
Abvt & 23N/62E & W1 & 833 & Greedy-CPU & 64 & 64 & 48.4 & 48.4 & 31.3 & 103.9 & 167.2 \\
Abvt & 23N/62E & W2 & 831 & RQ-SAFE & 64 & 64 & 71.9 & 71.9 & 22.9 & 77.6 & 691.9 \\
Abvt & 23N/62E & W2 & 831 & GNN-DAG & 64 & 64 & 71.9 & 71.9 & 22.8 & 77.3 & 590.9 \\
Abvt & 23N/62E & W2 & 831 & Latency-Aware & 64 & 64 & 65.6 & 65.6 & 39.8 & 94.7 & 179.6 \\
Abvt & 23N/62E & W2 & 831 & DRL-DDQN & 64 & 64 & 57.8 & 57.8 & 24.0 & 97.7 & 338.5 \\
Abvt & 23N/62E & W2 & 831 & Greedy-CPU & 64 & 64 & 46.9 & 46.9 & 31.3 & 102.2 & 181.6 \\
Abvt & 23N/62E & W2 & 832 & RQ-SAFE & 64 & 64 & 71.9 & 71.9 & 22.9 & 77.6 & 728.6 \\
Abvt & 23N/62E & W2 & 832 & GNN-DAG & 64 & 64 & 73.4 & 73.4 & 22.9 & 75.9 & 590.9 \\
Abvt & 23N/62E & W2 & 832 & Latency-Aware & 64 & 64 & 67.2 & 67.2 & 40.4 & 96.0 & 147.5 \\
Abvt & 23N/62E & W2 & 832 & DRL-DDQN & 64 & 64 & 57.8 & 57.8 & 24.0 & 97.6 & 288.8 \\
Abvt & 23N/62E & W2 & 832 & Greedy-CPU & 64 & 64 & 46.9 & 46.9 & 31.3 & 102.3 & 145.2 \\
Abvt & 23N/62E & W2 & 833 & RQ-SAFE & 64 & 64 & 71.9 & 71.9 & 22.9 & 75.7 & 658.7 \\
Abvt & 23N/62E & W2 & 833 & GNN-DAG & 64 & 64 & 73.4 & 73.4 & 22.9 & 75.9 & 601.1 \\
Abvt & 23N/62E & W2 & 833 & Latency-Aware & 64 & 64 & 65.6 & 65.6 & 39.8 & 94.7 & 170.5 \\
Abvt & 23N/62E & W2 & 833 & DRL-DDQN & 64 & 64 & 59.4 & 59.4 & 24.1 & 97.7 & 327.4 \\
Abvt & 23N/62E & W2 & 833 & Greedy-CPU & 64 & 64 & 46.9 & 46.9 & 31.3 & 102.3 & 177.0 \\
Abvt & 23N/62E & W3 & 831 & RQ-SAFE & 70 & 70 & 80.0 & 80.0 & 20.6 & 71.3 & 670.7 \\
Abvt & 23N/62E & W3 & 831 & GNN-DAG & 70 & 70 & 75.7 & 75.7 & 20.3 & 71.6 & 626.8 \\
Abvt & 23N/62E & W3 & 831 & Latency-Aware & 70 & 70 & 71.4 & 71.4 & 38.7 & 85.5 & 151.0 \\
Abvt & 23N/62E & W3 & 831 & DRL-DDQN & 70 & 70 & 64.3 & 64.3 & 21.9 & 89.2 & 284.7 \\
Abvt & 23N/62E & W3 & 831 & Greedy-CPU & 70 & 70 & 54.3 & 54.3 & 32.8 & 101.9 & 159.7 \\
Abvt & 23N/62E & W3 & 832 & RQ-SAFE & 70 & 70 & 78.6 & 78.6 & 20.6 & 71.5 & 656.3 \\
Abvt & 23N/62E & W3 & 832 & GNN-DAG & 70 & 70 & 75.7 & 75.7 & 20.3 & 71.7 & 613.2 \\
Abvt & 23N/62E & W3 & 832 & Latency-Aware & 70 & 70 & 71.4 & 71.4 & 38.7 & 85.5 & 148.1 \\
Abvt & 23N/62E & W3 & 832 & DRL-DDQN & 70 & 70 & 65.7 & 65.7 & 21.9 & 89.1 & 273.4 \\
Abvt & 23N/62E & W3 & 832 & Greedy-CPU & 70 & 70 & 55.7 & 55.7 & 33.0 & 102.1 & 142.3 \\
Abvt & 23N/62E & W3 & 833 & RQ-SAFE & 70 & 70 & 75.7 & 75.7 & 20.5 & 72.6 & 669.4 \\
Abvt & 23N/62E & W3 & 833 & GNN-DAG & 70 & 70 & 74.3 & 74.3 & 20.3 & 73.2 & 691.0 \\
Abvt & 23N/62E & W3 & 833 & Latency-Aware & 70 & 70 & 70.0 & 70.0 & 38.6 & 85.4 & 144.4 \\
Abvt & 23N/62E & W3 & 833 & DRL-DDQN & 70 & 70 & 65.7 & 65.7 & 21.8 & 89.0 & 317.7 \\
Abvt & 23N/62E & W3 & 833 & Greedy-CPU & 70 & 70 & 54.3 & 54.3 & 32.9 & 101.9 & 160.3 \\
Abvt & 23N/62E & W4 & 831 & RQ-SAFE & 70 & 70 & 87.1 & 87.1 & 21.1 & 72.6 & 621.9 \\
Abvt & 23N/62E & W4 & 831 & GNN-DAG & 70 & 70 & 84.3 & 84.3 & 20.9 & 72.5 & 686.7 \\
Abvt & 23N/62E & W4 & 831 & Latency-Aware & 70 & 70 & 68.6 & 68.6 & 37.7 & 86.2 & 166.7 \\
Abvt & 23N/62E & W4 & 831 & DRL-DDQN & 70 & 70 & 67.1 & 67.1 & 22.5 & 92.7 & 345.0 \\
Abvt & 23N/62E & W4 & 831 & Greedy-CPU & 70 & 70 & 48.6 & 48.6 & 30.6 & 100.8 & 142.6 \\
Abvt & 23N/62E & W4 & 832 & RQ-SAFE & 70 & 70 & 87.1 & 87.1 & 21.0 & 72.5 & 807.0 \\
Abvt & 23N/62E & W4 & 832 & GNN-DAG & 70 & 70 & 81.4 & 81.4 & 20.9 & 72.8 & 746.8 \\
Abvt & 23N/62E & W4 & 832 & Latency-Aware & 70 & 70 & 68.6 & 68.6 & 37.7 & 86.3 & 193.4 \\
Abvt & 23N/62E & W4 & 832 & DRL-DDQN & 70 & 70 & 65.7 & 65.7 & 22.4 & 92.5 & 360.5 \\
Abvt & 23N/62E & W4 & 832 & Greedy-CPU & 70 & 70 & 48.6 & 48.6 & 30.6 & 100.8 & 178.7 \\
Abvt & 23N/62E & W4 & 833 & RQ-SAFE & 70 & 70 & 87.1 & 87.1 & 21.0 & 72.6 & 737.2 \\
Abvt & 23N/62E & W4 & 833 & GNN-DAG & 70 & 70 & 82.9 & 82.9 & 20.9 & 73.3 & 728.6 \\
Abvt & 23N/62E & W4 & 833 & Latency-Aware & 70 & 70 & 68.6 & 68.6 & 37.7 & 86.3 & 159.4 \\
Abvt & 23N/62E & W4 & 833 & DRL-DDQN & 70 & 70 & 65.7 & 65.7 & 22.2 & 92.1 & 301.3 \\
Abvt & 23N/62E & W4 & 833 & Greedy-CPU & 70 & 70 & 47.1 & 47.1 & 30.4 & 100.4 & 162.1 \\
\addlinespace[1.2pt]
Aconet & 23N/62E & W1 & 831 & RQ-SAFE & 64 & 64 & 71.9 & 71.9 & 23.4 & 77.3 & 653.1 \\
Aconet & 23N/62E & W1 & 831 & GNN-DAG & 64 & 64 & 71.9 & 71.9 & 23.4 & 75.9 & 688.1 \\
Aconet & 23N/62E & W1 & 831 & Latency-Aware & 64 & 64 & 59.4 & 59.4 & 39.9 & 95.9 & 153.1 \\
Aconet & 23N/62E & W1 & 831 & DRL-DDQN & 64 & 64 & 64.1 & 64.1 & 24.9 & 94.1 & 291.4 \\
Aconet & 23N/62E & W1 & 831 & Greedy-CPU & 64 & 64 & 48.4 & 48.4 & 31.3 & 106.0 & 149.4 \\
Aconet & 23N/62E & W1 & 832 & RQ-SAFE & 64 & 64 & 71.9 & 71.9 & 23.4 & 77.5 & 662.4 \\
Aconet & 23N/62E & W1 & 832 & GNN-DAG & 64 & 64 & 70.3 & 70.3 & 23.3 & 78.0 & 705.1 \\
Aconet & 23N/62E & W1 & 832 & Latency-Aware & 64 & 64 & 59.4 & 59.4 & 39.9 & 95.9 & 146.9 \\
Aconet & 23N/62E & W1 & 832 & DRL-DDQN & 64 & 64 & 64.1 & 64.1 & 24.9 & 94.1 & 338.5 \\
Aconet & 23N/62E & W1 & 832 & Greedy-CPU & 64 & 64 & 48.4 & 48.4 & 31.3 & 106.0 & 136.8 \\
Aconet & 23N/62E & W1 & 833 & RQ-SAFE & 64 & 64 & 71.9 & 71.9 & 23.4 & 77.2 & 713.7 \\
Aconet & 23N/62E & W1 & 833 & GNN-DAG & 64 & 64 & 71.9 & 71.9 & 23.4 & 76.6 & 715.8 \\
Aconet & 23N/62E & W1 & 833 & Latency-Aware & 64 & 64 & 59.4 & 59.4 & 39.9 & 95.9 & 151.3 \\
Aconet & 23N/62E & W1 & 833 & DRL-DDQN & 64 & 64 & 64.1 & 64.1 & 24.9 & 94.1 & 327.6 \\
Aconet & 23N/62E & W1 & 833 & Greedy-CPU & 64 & 64 & 48.4 & 48.4 & 31.3 & 106.0 & 141.2 \\
Aconet & 23N/62E & W2 & 831 & RQ-SAFE & 64 & 64 & 71.9 & 71.9 & 22.7 & 75.4 & 629.4 \\
Aconet & 23N/62E & W2 & 831 & GNN-DAG & 64 & 64 & 71.9 & 71.9 & 22.6 & 75.3 & 682.4 \\
Aconet & 23N/62E & W2 & 831 & Latency-Aware & 64 & 64 & 65.6 & 65.6 & 39.8 & 94.7 & 156.0 \\
Aconet & 23N/62E & W2 & 831 & DRL-DDQN & 64 & 64 & 65.6 & 65.6 & 24.3 & 91.4 & 281.4 \\
Aconet & 23N/62E & W2 & 831 & Greedy-CPU & 64 & 64 & 46.9 & 46.9 & 31.3 & 104.2 & 140.8 \\
Aconet & 23N/62E & W2 & 832 & RQ-SAFE & 64 & 64 & 71.9 & 71.9 & 22.7 & 75.4 & 732.7 \\
Aconet & 23N/62E & W2 & 832 & GNN-DAG & 64 & 64 & 71.9 & 71.9 & 22.6 & 75.8 & 712.3 \\
Aconet & 23N/62E & W2 & 832 & Latency-Aware & 64 & 64 & 67.2 & 67.2 & 40.4 & 96.0 & 157.4 \\
Aconet & 23N/62E & W2 & 832 & DRL-DDQN & 64 & 64 & 65.6 & 65.6 & 24.3 & 91.4 & 307.7 \\
Aconet & 23N/62E & W2 & 832 & Greedy-CPU & 64 & 64 & 46.9 & 46.9 & 31.3 & 104.3 & 147.9 \\
Aconet & 23N/62E & W2 & 833 & RQ-SAFE & 64 & 64 & 71.9 & 71.9 & 22.7 & 75.4 & 726.8 \\
Aconet & 23N/62E & W2 & 833 & GNN-DAG & 64 & 64 & 71.9 & 71.9 & 22.6 & 75.4 & 700.7 \\
Aconet & 23N/62E & W2 & 833 & Latency-Aware & 64 & 64 & 65.6 & 65.6 & 39.8 & 94.7 & 143.2 \\
Aconet & 23N/62E & W2 & 833 & DRL-DDQN & 64 & 64 & 65.6 & 65.6 & 24.3 & 91.3 & 282.8 \\
Aconet & 23N/62E & W2 & 833 & Greedy-CPU & 64 & 64 & 46.9 & 46.9 & 31.3 & 104.3 & 145.2 \\
Aconet & 23N/62E & W3 & 831 & RQ-SAFE & 70 & 70 & 84.3 & 84.3 & 20.4 & 69.5 & 627.1 \\
Aconet & 23N/62E & W3 & 831 & GNN-DAG & 70 & 70 & 87.1 & 87.1 & 20.1 & 68.8 & 742.2 \\
Aconet & 23N/62E & W3 & 831 & Latency-Aware & 70 & 70 & 71.4 & 71.4 & 38.7 & 85.5 & 157.9 \\
Aconet & 23N/62E & W3 & 831 & DRL-DDQN & 70 & 70 & 74.3 & 74.3 & 22.3 & 83.8 & 313.5 \\
Aconet & 23N/62E & W3 & 831 & Greedy-CPU & 70 & 70 & 54.3 & 54.3 & 32.8 & 102.8 & 141.1 \\
Aconet & 23N/62E & W3 & 832 & RQ-SAFE & 70 & 70 & 84.3 & 84.3 & 20.4 & 69.4 & 685.8 \\
Aconet & 23N/62E & W3 & 832 & GNN-DAG & 70 & 70 & 87.1 & 87.1 & 20.1 & 68.9 & 765.7 \\
Aconet & 23N/62E & W3 & 832 & Latency-Aware & 70 & 70 & 71.4 & 71.4 & 38.7 & 85.5 & 172.9 \\
Aconet & 23N/62E & W3 & 832 & DRL-DDQN & 70 & 70 & 75.7 & 75.7 & 22.3 & 83.8 & 309.7 \\
Aconet & 23N/62E & W3 & 832 & Greedy-CPU & 70 & 70 & 55.7 & 55.7 & 33.0 & 103.1 & 173.1 \\
Aconet & 23N/62E & W3 & 833 & RQ-SAFE & 70 & 70 & 84.3 & 84.3 & 20.4 & 69.2 & 667.4 \\
Aconet & 23N/62E & W3 & 833 & GNN-DAG & 70 & 70 & 87.1 & 87.1 & 20.1 & 69.0 & 661.4 \\
Aconet & 23N/62E & W3 & 833 & Latency-Aware & 70 & 70 & 70.0 & 70.0 & 38.6 & 85.4 & 157.9 \\
Aconet & 23N/62E & W3 & 833 & DRL-DDQN & 70 & 70 & 72.9 & 72.9 & 22.1 & 83.5 & 282.3 \\
Aconet & 23N/62E & W3 & 833 & Greedy-CPU & 70 & 70 & 54.3 & 54.3 & 32.9 & 102.9 & 145.3 \\
Aconet & 23N/62E & W4 & 831 & RQ-SAFE & 70 & 70 & 85.7 & 85.7 & 21.0 & 71.0 & 624.2 \\
Aconet & 23N/62E & W4 & 831 & GNN-DAG & 70 & 70 & 88.6 & 88.6 & 20.7 & 71.5 & 651.3 \\
Aconet & 23N/62E & W4 & 831 & Latency-Aware & 70 & 70 & 68.6 & 68.6 & 37.7 & 86.2 & 149.0 \\
Aconet & 23N/62E & W4 & 831 & DRL-DDQN & 70 & 70 & 70.0 & 70.0 & 22.5 & 86.4 & 300.3 \\
Aconet & 23N/62E & W4 & 831 & Greedy-CPU & 70 & 70 & 48.6 & 48.6 & 30.6 & 102.0 & 144.9 \\
Aconet & 23N/62E & W4 & 832 & RQ-SAFE & 70 & 70 & 85.7 & 85.7 & 21.0 & 71.0 & 692.2 \\
Aconet & 23N/62E & W4 & 832 & GNN-DAG & 70 & 70 & 90.0 & 90.0 & 20.7 & 71.3 & 653.6 \\
Aconet & 23N/62E & W4 & 832 & Latency-Aware & 70 & 70 & 68.6 & 68.6 & 37.7 & 86.3 & 151.1 \\
Aconet & 23N/62E & W4 & 832 & DRL-DDQN & 70 & 70 & 70.0 & 70.0 & 22.5 & 86.4 & 296.0 \\
Aconet & 23N/62E & W4 & 832 & Greedy-CPU & 70 & 70 & 48.6 & 48.6 & 30.6 & 102.0 & 149.4 \\
Aconet & 23N/62E & W4 & 833 & RQ-SAFE & 70 & 70 & 85.7 & 85.7 & 21.0 & 71.2 & 659.3 \\
Aconet & 23N/62E & W4 & 833 & GNN-DAG & 70 & 70 & 87.1 & 87.1 & 20.7 & 71.7 & 649.3 \\
Aconet & 23N/62E & W4 & 833 & Latency-Aware & 70 & 70 & 68.6 & 68.6 & 37.7 & 86.3 & 165.0 \\
Aconet & 23N/62E & W4 & 833 & DRL-DDQN & 70 & 70 & 70.0 & 70.0 & 22.3 & 86.0 & 296.2 \\
Aconet & 23N/62E & W4 & 833 & Greedy-CPU & 70 & 70 & 47.1 & 47.1 & 30.4 & 101.6 & 154.2 \\
\end{longtable}
\end{ThreePartTable}
\clearpage

\section{Scope Boundaries and Deployment Extensions}
\label{app:scope_boundaries}
This appendix section collects the main deployment-scope and implementation boundaries so that the main text can focus on the orchestration design and experimental findings. The reported experiments are designed to evaluate validation-controlled online SFC-DAG orchestration under matched topology, workload, seed, and service-pool states. The following points clarify how the framework can be extended in broader deployments.

\subsection{Queueing and Traffic Models}
The reported scheduler uses the observable FCFS proxy $\widehat{W}_s(t)=Q_s(t)/\mu_s$ as a lightweight online estimator. The robustness analysis and Appendix the main-text queue-error robustness figure evaluate bounded queue-estimation error and validation-side margins. A complete queueing-model comparison under non-FCFS scheduling, heavy-tailed service times, background interference, and adversarial burst patterns would require a separate runtime study. In deployment, the margin $\epsilon_q(p)$ can be calibrated from recent validation residuals, as described in Appendix~\ref{app:implementation_scope}.

\subsection{Resource Reuse and Activation Calibration}
The effective resource increment $\Delta r^r_{v,s}$ is computed from benchmark VNF demand, cold/warm instance status, and conservative reuse rules. This makes the reported matched replay reproducible. In a production system, the reuse coefficient $\phi_{v,s}(t)$ should be calibrated from telemetry, for example by measuring marginal CPU usage after assigning additional tasks to warm instances. Such calibration changes the validation ledger, while preserving the separation between learned candidate ordering and final feasibility validation.

\subsection{Commutable Semantics and New VNF Types}
The semantic compatibility matrix is intentionally conservative. New VNF types can be introduced by declaring their type, statefulness, side-effect class, and admissible commutable partners. Unknown pairs are kept in their original dependency order until this metadata is available. A future extension can combine rule-based safety with learned semantic compatibility, provided that dependency preservation and validation remain explicit.

\subsection{Baselines, Testbeds, and Larger Deployments}
The evaluation compares representative heuristic, latency-aware, DRL, and graph-aware baselines under matched traces. It is intended to position the proposed request--resource joint-scheduling pipeline rather than to cover every hybrid learning-and-verification implementation. Future work will add real edge testbed measurements, container start-up effects, cloud-edge or multi-domain orchestration, and integrated migration, autoscaling, and failure recovery. These extensions affect deployment realism and long-horizon resource governance, while the current paper focuses on per-request online SFC-DAG orchestration with explicit request-local completion checks and traceable state updates.